\documentclass[journal]{IEEEtran}
\usepackage{algorithm, algpseudocode}
\usepackage{authblk, cite, enumitem, graphicx, hyperref, pgfplots, subfig}
\pgfplotsset{compat=1.18}
\usepackage{amsfonts, amsmath, amssymb, amsthm}
\usepackage{booktabs, tabularx}  %
\usepackage{CJKutf8, dsfont}
\usepackage{bm}
\usepackage[normalem]{ulem}
\newlist{todolist}{itemize}{2}
\setlist[todolist]{label=$\square$}
\hypersetup{colorlinks=true, citecolor=blue(ryb), linkcolor=blue(ryb), urlcolor=cyan}
\PassOptionsToPackage{hyphens}{url}
\definecolor{azure}{rgb}{0.0, 0.5, 1.0}
\definecolor{frenchblue}{rgb}{0.0, 0.45, 0.73}
\definecolor{forestgreen(traditional)}{rgb}{0.0, 0.27, 0.13}
\definecolor{mygreen}{rgb}{0.09, 0.45, 0.27}
\definecolor{myblue}{rgb}{0.2383,0.5195,0.7734}
\definecolor{mygreen}{rgb}{0.6445,0.9297,0.0039}
\definecolor{darklavender}{rgb}{0.45, 0.31, 0.59}
\definecolor{americanrose}{rgb}{1.0, 0.01, 0.24}
\definecolor{pigblue}{rgb}{0.2, 0.2, 0.6}
\definecolor{blue(ryb)}{rgb}{0.01, 0.28, 1.0}
\definecolor{amethyst}{rgb}{0.6, 0.4, 0.8}
\definecolor{deepmagenta}{rgb}{0.8, 0.0, 0.8}
\definecolor{carminered}{rgb}{1.0, 0.0, 0.22}
\definecolor{iris}{rgb}{0.35, 0.31, 0.81}

\newtheorem{proposition}{\hspace{0pt}\bf Proposition}

\newcommand*{\herm}{{\mkern-1.5mu\mathsf{H}}}

\DeclareMathOperator*{\argmin}{arg\,min}

\def \tr      {\text{\normalfont tr}   }

\renewcommand \Re    [1] {\text{\normalfont Re} #1}

\newcommand{\minimize}{\operatornamewithlimits{minimize}}
\newcommand{\st}{\operatornamewithlimits{s.t.}}

\def\ccal0{{\ensuremath{\mathcal 0}}}

\def\bb0{{\ensuremath{\boldsymbol 0}}}

\definecolor{crimson}{rgb}{0.86, 0.08, 0.24}
\definecolor{scarlet}{rgb}{1.0, 0.13, 0.0}
\definecolor{hookersgreen}{rgb}{0.0, 0.44, 0.0}
\definecolor{cultramarine}{rgb}{0.25, 0.0, 1.0}

\newcommand\revised[1] {#1}

\title{A Majorization-Minimization Framework for Activity Detection in Mixed Near-and-Far-Field Random Access}
\author{Xinjue~Wang,~\IEEEmembership{Member,~IEEE},
        Zhi-Yong~Wang,~\IEEEmembership{Member,~IEEE},
        Sergiy~A.~Vorobyov,~\IEEEmembership{Fellow,~IEEE},
        Esa~Ollila,~\IEEEmembership{Senior Member,~IEEE},
        Gayan~Amarasuriya~Aruma~Baduge,~\IEEEmembership{Senior Member,~IEEE}
        and~Mojtaba~Vaezi,~\IEEEmembership{Senior Member,~IEEE}%
\thanks{Xinjue Wang, Sergiy Vorobyov, and Esa Ollila are with the Department of Information and Communications Engineering, Aalto University, 02150 Espoo, Finland (e-mail: xjw@ieee.org; sergiy.vorobyov@aalto.fi; esa.ollila@aalto.fi). Their work was supported by the Research Council of Finland under Grant 359848. \textit{(Corresponding authors: Xinjue Wang and Sergiy A. Vorobyov.)}}%
\thanks{Zhi-Yong Wang is with the State Key Laboratory of Ocean Sensing, Ocean College \& ZJU-Hangzhou Global Scientific and Technological Innovation Center, Zhejiang University, Hangzhou 310000, China (e-mail: z.y.wang@my.cityu.edu.hk).}%
\thanks{Gayan Amarasuriya Aruma Baduge is with the School of ECBE, Southern Illinois University, Carbondale, IL 62918 USA (e-mail: gayan.baduge@siu.edu). His work in part  has been  supported by the NSF under
Grant CCF-2326621.}%
\thanks{Mojtaba Vaezi is with the ECE Department, Villanova University, Villanova, PA 19085 USA (e-mail: mvaezi@villanova.edu). His work was supported by the U.S. National
Science Foundation  under Grant CCF-2326622.} }

\begin{document}
\maketitle
\begin{abstract}
Grant-free random access in massive machine-type communications requires detecting a small active set from length-$L$ uplink pilots received at an $M$-antenna base station. 
Classical covariance-based detectors are largely built on the far-field (FF) model, where the $M$ antenna-domain observations are treated as independent snapshots, and inference reduces to an $L\times L$ covariance description. 
This model becomes inadequate in mixed near-field (NF) and FF access, where NF devices induce device-specific structured spatial covariances and the aggregate observation no longer fits the FF snapshot structure. 
In this paper, we develop a unified covariance-aware Rician framework for activity detection that treats FF and NF devices by a single likelihood model. 
Within this framework, we propose a majorization-minimization projected gradient descent (MM-PGD) detector for the resulting relaxed likelihood. 
For scalable exact likelihood evaluation, we further derive a Kronecker--Woodbury implementation that exploits the mixed NF/FF covariance structure, and avoids factorizing the full $LM\times LM$ covariance matrix.
Numerical results show that MM-PGD is on par with the strongest coordinate-wise baseline in the all-FF limit, while its advantage becomes more pronounced as the fraction of NF users increases under the tested regimes.
\end{abstract}

\begin{IEEEkeywords}
Grant-free random access, activity detection, mixed near-/far-field channels, covariance-aware inference, majorization-minimization.
\end{IEEEkeywords}

\section{Introduction}
\label{sec:intro}

\IEEEPARstart{G}{rant-free} random access is a core access mechanism for massive machine-type communications (mMTC), where only a small subset of devices is active in each coherence block~\cite{bockelmann2018towards,chen2020massive,dawy2016toward,shahab2020grant}.
Activity detection (AD) aims to recover this active set from uplink pilots and directly affects both access reliability and downstream data detection~\cite{liu2018sparse,liu2018massiveAMP}.
The problem becomes challenging when the pilot dimension is much smaller than the number of registered devices, so the base station (BS) must infer a sparse activity pattern from a noisy superposition of user signatures.
With large arrays, this task is further complicated in mixed near-/far-field (mixed NF/FF) access~\cite{Liu2023NFReview,Lu2024NFXLMIMOTutorial,Liu2025NFComprehensive}, where a non-negligible part of the device population may operate in the radiating near-field (NF) while other devices remain effectively far-field (FF).
In this paper, we focus on AD in this mixed NF/FF setting at a multiple-input multiple-output (MIMO) BS with $L$ pilot symbols and $M$ antennas.

Most covariance-based AD methods are developed for FF channels under an uncorrelated Rayleigh fading model~\cite{liu2018sparse,chen2018sparse,fengler2021nonTIT}.
Under this model, the antenna-domain observations can be treated as independent snapshots, and inference is carried out through an $L\times L$ sample covariance~\cite{liu2018sparse,chen2018sparse}.
This snapshot structure underlies a broad class of FF detectors, including approximate message passing (AMP) variants~\cite{donoho2009AMP,liu2018massiveAMP}, sparse Bayesian learning (SBL) methods~\cite{tipping2001sparsetruSBL,wipf2007empiricalMSBL}, coordinate-wise covariance learning (CWO)~\cite{haghighatshoar2018improved,fengler2021nonTIT}, and covariance-based matching pursuit (CL-MP)~\cite{marata2024activity}.
For covariance learning-type methods, this FF snapshot approximation leads to rank-one covariance updates and simple inverse recursions. It also keeps covariance learning in the reduced $L\times L$ domain, because the antenna-domain observations can be treated as repeated snapshots.

This existing FF snapshot structure in ~\cite{liu2018sparse,chen2018sparse,fengler2021nonTIT}  becomes inaccurate in mixed NF/FF access. 
As the array aperture grows, a non-negligible fraction of devices may lie in the radiating NF~\cite{Liu2023NFReview,Lu2024NFXLMIMOTutorial,Liu2025NFComprehensive}. 
NF propagation introduces spherical wavefronts, user-specific spatial covariance, and antenna-domain coupling~\cite{Lu2024NFXLMIMOTutorial}. 
These effects are absent in the FF model, where each device covariance is modeled as isotropic, i.e., as a scaled identity matrix under the uncorrelated fading channel model. 
When NF and FF users coexist, the columns of the received pilot matrix are no longer independent across antennas. 
The AD problem therefore departs from the classical $L\times L$ snapshot covariance description and has to be written in a joint $LM$-dimensional model with device-dependent covariance structure. 
Consequently, the isotropic approximation becomes inadequate for the channel statistics, and the rank-one covariance updates that make FF coordinate-wise solvers efficient are no longer available.

This paper considers AD in mixed NF/FF access beyond the classical FF snapshot setting. 
We formulate AD under a unified Rician covariance-aware likelihood on the full $LM$-dimensional covariance domain, with FF and NF users modeled by isotropic and user-dependent structured spatial covariances, respectively. 
We then develop a majorization-minimization projected gradient descent (MM-PGD) algorithm with joint activity updates under the relaxed likelihood. 
For scalable exact likelihood evaluation, we further derive a Kronecker--Woodbury implementation that fully exploits the mixed NF/FF covariance structure.

\subsection{Related Work}
The classical FF snapshot model underlies both multiple-measurement-vector (MMV)-type AMP and SBL formulations~\cite{donoho2009AMP,liu2018massiveAMP,tipping2001sparsetruSBL,wipf2007empiricalMSBL} and covariance-based detectors~\cite{haghighatshoar2018improved,fengler2021nonTIT,ollila2024matching,marata2024activity,chen2021phase}. 
The covariance-based detectors have also been extended to more general communication settings, including robust activity detection~\cite{wang2025ADicassp,wang2025RobustADTSP}, joint activity/data detection~\cite{wang2021efficientJoint,zhang2023jointBiGAMP}, cooperative multi-cell processing~\cite{chen2019covarianceICC,wang24MultiCellAD}, and cell-free architectures~\cite{zhang24ADCELLFREESBL,rajoriya2024novelCELLFREE,ganesan2021clustering}. 
These formulations still inherit the FF modeling structure that makes reduced-dimensional covariance learning and low-rank coordinate updates effective.
\revised{Compressed sensing (CS) methods for AD, including AMP and generalized approximate message passing (GAMP), recover the instantaneous channel matrix and infer device activity from its row support~\cite{liu2018massiveAMP,senel2018grant,zhang2023jointBiGAMP}. By contrast, covariance-based methods marginalize the instantaneous channels and estimate the activity variables directly from stored first- and second-order channel statistics~\cite{chen2018sparse,fengler2021nonTIT}. In the mixed NF/FF setting, the nonzero Rician means and user-dependent cross-antenna covariances couple the vectorized observation and are not represented by the classical FF AMP and GAMP formulations; accommodating them would require model-specific priors and denoisers~\cite{wangcas2025TWCNearField}.}

Among the prior directions discussed above, the one most directly connected to the present paper is the Rician maximum-likelihood estimation (MLE) formulation for AD \cite{Tian2022UnsourcedRician, Liu2024CWOMMLE, wangcas2025TWCNearField}. 
An earlier work~\cite{Tian2022UnsourcedRician} showed that explicit line-of-sight (LoS) mean vectors should be incorporated into likelihood-based random-access detection. 
Later,~\cite{Liu2024CWOMMLE} extended this formulation to grant-free AD under an isotropic covariance model.
Under this model, the problem remains in the $L\times L$ covariance domain and coordinate-wise rank-one updates are still available. 
Recent work in~\cite{wangcas2025TWCNearField} extends this Rician-MLE from isotropic FF covariance to structured low-rank NF spatial covariance.
\revised{Its inexact coordinate-descent (CD) solver is referred to as NF-CD in this paper.}
Retaining structured covariance shifts the likelihood to the vectorized $LM$-dimensional observation space, where the rank-one update structure disappears while the solver remains coordinate-wise.
Although the system model there includes both NF and FF devices, the detection-performance analysis still treats them as separate cases.
Taken together, this leaves open how to treat structured NF covariance and FF devices within one unified mixed-field mechanism.

The works \cite{Tian2022UnsourcedRician, Liu2024CWOMMLE, wangcas2025TWCNearField} are based on explicit LoS modeling and structured NF covariance, but do not develop a unified mixed NF/FF detection framework.
The remaining problem is not only how to model device covariance more faithfully. 
At the system level, mixed NF/FF access should be treated as one detection problem.
The present paper is motivated by this unified viewpoint and treats FF devices, NF devices, and their coexistence within one covariance-aware detection framework.

\subsection{Contributions}
This paper develops a covariance-aware Rician framework for mixed NF/FF activity detection.
The framework treats FF devices, NF devices, and their coexistence within one activity-detection model. The FF-only and NF-only settings then follow as covariance-induced specializations of the same framework.

Under this viewpoint, we formulate mixed NF/FF activity detection as a relaxed MLE problem in the vectorized observation domain. The formulation retains both the LoS mean %
and the device-specific spatial covariance, %
and thereby places isotropic FF covariance, structured NF covariance, and their coexistence within one statistical model. 
We further derive the exact gradient of the relaxed likelihood and show that the resulting mixed NF/FF problem is full-vector and globally coupled, in the sense that all activity variables are updated jointly.

This  covariance structural characterization clarifies why the classical FF rank-one coordinate-update logic does not extend naturally to mixed NF/FF detection. We therefore develop MM-PGD as a generic full-vector update framework whose basic mechanism remains the same across FF, NF, and mixed NF/FF regimes. Here MM-PGD is naturally motivated by the unified mixed-field likelihood. Combined with a backtracking step-size rule, the resulting algorithm admits standard monotone-descent and projected-stationarity guarantees.

We further derive an exact Kronecker--Woodbury realization for efficient exact evaluation of the mixed NF/FF detector. The key is to exploit the pilot-induced Kronecker structure jointly with the low-rank NF covariance factors. \revised{This structural decomposition, formalized in Proposition~\ref{prop:woodbury_decomp}, is the novelty that makes the Kronecker--Woodbury evaluation applicable to the mixed NF/FF covariance.} This reduces the dominant $LM$-dimensional covariance factorization to smaller factors whose dimensions are determined by the FF/NF composition. In the all-FF case, the resulting realization recovers the classical $L\times L$ covariance-domain evaluation. In the mixed FF/NF case, it is computationally more favorable than structured coordinate-wise NF realizations. In the all-NF case, it remains exact under the same factorization mechanism. Runtime results validate the corresponding wall-clock savings. In terms of detection performance, MM-PGD remains competitive with the strongest isotropic Rician MLE baseline in the all-FF case and achieves lower miss-detection rates as the fraction of NF users increases.

\subsection{Organization}
The remainder of this paper is organized as follows.
Section~\ref{sec:sysmd} introduces the mixed NF/FF system model.
Section~\ref{sec:prob_form} formulates the vectorized likelihood model and the covariance-aware relaxed MLE problem.
Section~\ref{sec:mm_alg} develops the MM-PGD framework together with its exact Kronecker--Woodbury realization.
Section~\ref{sec:exp_setup} reports numerical experiments.
Section~\ref{sec:conclusion} concludes the paper.
Technical proofs are presented in  the appendices.

\subsection{Notation}
Boldface lowercase letters denote vectors and boldface uppercase letters denote matrices. The operators $(\cdot)^\top$, $(\cdot)^\herm$, and $(\cdot)^\ast$ denote transpose, Hermitian transpose, and complex conjugation, respectively. The identity matrix is denoted by $\bm I$, while $\mathrm{vec}(\cdot)$ and $\tr(\cdot)$ denote vectorization and trace. The determinant and inverse of matrix $\bm A$ are denoted as $| \bm A |$ and $\bm A^{-1}$, respectively. The notation $\otimes$ stands for the Kronecker product. The complex number set is denoted by $\mathbb{C}$, $|\mathcal{K}|$ stands for the cardinality of the set $\mathcal{K}$. The function $\mathbb{I}\{a\ge\tau\}$ takes value 1 if the condition in the argument is satisfied and 0 otherwise, while $\Pi_{[0,1]^N}$ denotes element-wise clipping onto $[0,1]$ and $I_{[0,1]^N}$ stands for the indicator function of the set $[0,1]^N$, \revised{equal to $0$ on $[0,1]^N$ and $+\infty$ otherwise}. Complex Gaussian distribution with mean $m$ and variance $\sigma^2$ is denoted as $\mathcal{CN}(m, \sigma^2)$.

\section{System Model} \label{sec:sysmd}

We consider an uplink single-cell massive MIMO system designed for  mMTC. 
The BS is equipped with a uniform linear array (ULA) deployed along the $y$-axis, centered at the origin. 
The array consists of $M$ antenna elements with half-wavelength spacing $d = \lambda/2$, where $\lambda$ denotes the carrier wavelength. 
The spatial coordinate of the $m$-th antenna element is denoted by $\bm{u}_m = \left[ 0, \; (m - \frac{M+1}{2})d \right]^\top$ for $m=1, \dots, M$, and $\bm{u}_m \in \mathbb{R}^2$.

The network serves $N$ single-antenna devices randomly distributed in the coverage area. 
We consider a sporadic traffic scenario where only a small subset of devices, denoted by the set $\mathcal{K} \subset \{1, \dots, N\}$, are active within a given coherence interval. 
The number of active devices is $K = |\mathcal{K}|$, satisfying $K \ll N$.
\revised{In line with prior covariance-based formulations~\cite{chen2018sparse,fengler2021nonTIT,Tian2022UnsourcedRician,Liu2024CWOMMLE,wangcas2025TWCNearField}, the BS is assumed to have access only to the \emph{long-term statistical channel state information (CSI)} of the registered device pool. The instantaneous channel realizations are not required. The online task is to infer the activity vector from the current pilot observation.}

\subsection{Signal Transmission Model}

Each device $n$ is pre-assigned a unique signature sequence (pilot) $\bm{s}_n \in \mathbb{C}^{L}$ of length $L$, with unit norm constraint $\|\bm{s}_n\|_2 = 1$. Let $\alpha_n \in \{0, 1\}$ denote the binary activity indicator for the $n$-th device:
\begin{equation}
    \alpha_n = 
    \begin{cases} 
    1, & \text{if } n \in \mathcal{K}, \\
    0, & \text{otherwise.}
    \end{cases}
\end{equation}
The received signal matrix $\bm{Y} \in \mathbb{C}^{L \times M}$ at the BS is the superposition of signals from all active devices, modeled as:
\begin{equation}
    \bm{Y} = \sum_{n=1}^{N} \alpha_n \bm{s}_n \bm{h}_n^\top + \bm{W},
    \label{eq:received_signal}
\end{equation}
where $\bm{h}_n \in \mathbb{C}^{M}$ represents the uplink channel vector for the $n$-th device, and $\bm{W} \in \mathbb{C}^{L \times M}$ denotes the additive white Gaussian noise matrix with independent and identically distributed (i.i.d.) entries following $\mathcal{CN}(0, \sigma_w^2)$ with $\sigma_w^2$ standing for the noise power.

\subsection{Mixed Near-/Far-Field Channel Model}

We consider a mixed scenario where some devices are located in the radiating NF region of the BS array, while other devices are located in the FF region.
The Rayleigh distance is given by $Z_{\text{Ray}} = 2D^2 / \lambda$, where $D$ is the array aperture, and $D=(M-1)d$ for ULA.
In this work, NF refers to the \emph{radiating NF} (RNF) regime~\cite{Liu2023NFReview, Lu2024NFXLMIMOTutorial}, for which we consider the distance range\footnote{The region $0<r\le r_{\text{RNF}}$ is commonly referred to as the \emph{reactive NF}, where evanescent/reactive field components may dominate and the simple spherical-wave propagation model becomes less accurate~\cite{Liu2023NFReview}. 
Following the common practice in NF communication studies, we focus on the radiating NF and exclude the reactive NF in this paper.}
\begin{equation}
    r_{\text{RNF}} < r < Z_{\text{Ray}}, \qquad r_{\text{RNF}} \triangleq 0.62\sqrt{\frac{D^3}{\lambda}} .
\end{equation}
We define the index sets as
\begin{align}
    \mathcal{N}_{\mathrm{NF}} & \triangleq \left\{ n \in \{1,\dots,N\} : r_{\text{RNF}} < r_n < Z_{\text{Ray}} \right\}, \\
    \mathcal{N}_{\mathrm{FF}} & \triangleq \{1,\dots,N\}\setminus \mathcal{N}_{\mathrm{NF}} .
\end{align}
The overall statistical channel model is unified as
\begin{equation}
    \bm h_n \sim \mathcal{CN}(\bar{\bm h}_n,\bm R_n), \qquad n=1,\dots,N,
    \label{eq:rician_unified}
\end{equation}
where the long-term channel statistics pair $(\bar{\bm h}_n,\bm R_n)$ is specified differently for NF and FF devices.
\revised{The proposed detector follows the covariance-based paradigm. The instantaneous channel $\bm h_n$ is random and is re-drawn across access attempts. Detection relies only on the stored statistics $(\bar{\bm h}_n,\bm R_n)$ and the current observation $\bm Y$. The realized $\bm h_n$ is not used. This differs from the compressed-sensing-based approach. There, the instantaneous channel matrix is the sparse signal that needs to be estimated, and the activity pattern is given by its support~\cite{liu2018massiveAMP,senel2018grant,zhang2023jointBiGAMP}.}
In the RNF, the plane-wave assumption no longer holds, and hence the channel must be characterized by using spherical wavefronts, as detailed below.

\subsubsection{Physical Multipath Decomposition for NF Devices}
For each NF device $n\in\mathcal{N}_{\mathrm{NF}}$, we define its location using polar coordinates $(r_n, \theta_n)$, where $r_n$ denotes the distance from the device to the center of the array, and $\theta_n \in [0, \pi]$ denotes the angle of arrival.
Let the location of a device with polar parameters $(r,\theta)$ be represented in Cartesian coordinates as
\begin{equation}
    \bm u_{\mathrm{user}}(r,\theta) \triangleq [r\cos\theta,\ r\sin\theta]^\top \in \mathbb{R}^2 .
\end{equation}

Let $\bm{b}(r, \theta) \in \mathbb{C}^M$ denote the steering vector. Based on the spherical-wave propagation model~\cite{Liu2023NFReview,Lu2024NFXLMIMOTutorial}, the phase difference depends on the specific Euclidean distance to each antenna. 
The $m$-th element of $\bm{b}(r, \theta)$ is given by
\begin{equation}
    [\bm{b}(r, \theta)]_m = \frac{1}{\sqrt{M}}
    \exp\left( -j \frac{2\pi}{\lambda} \left( d_m(r, \theta) - r \right) \right),
    \label{eq:steering_vec}
\end{equation}
where $d_m(r,\theta)\triangleq \|\bm u_{\mathrm{user}}(r,\theta)-\bm u_m\|_2$ is the Euclidean distance between the device and the $m$-th antenna element located at $\bm u_m=[0,\delta_m]^\top$ with $\delta_m = \left( m-\frac{M+1}{2} \right) d$.
For a ULA, this distance can be explicitly written as
\begin{equation}
    d_m(r, \theta) = \sqrt{r^2 + \delta_m^2 - 2 r \delta_m \sin\theta}.
    \label{eq:dist_exact}
\end{equation}

The channel $\bm{h}_n$ is composed of a deterministic LoS path and a superposition of $L_n$ non-line-of-sight (NLoS) scattering paths (clusters). The channel vector is expressed as
\begin{equation}
    \bm{h}_n = \underbrace{\beta_{n,0} \bm{b}(r_n, \theta_n)}_{\text{LoS Component}} + \underbrace{\sum_{\ell=1}^{L_n} \beta_{n,\ell} \bm{b}(r_{n,\ell}, \theta_{n,\ell})}_{\text{NLoS Component}},
    \label{eq:physical_channel}
\end{equation}
where $\beta_{n,0} \in \mathbb{C}$ is the complex gain of the LoS path determined by $(r_n, \theta_n)$. Similarly, $\beta_{n,\ell}$, $r_{n,\ell}$, and $\theta_{n,\ell}$ denote the complex gain, distance, and angle of the $\ell$-th scatterer, respectively.
These scatterer-level quantities serve as a physical parametrization of the long-term channel statistics; the detector itself uses the resulting channel statistics $(\bar{\bm h}_n,\bm R_n)$ during activity detection.

\subsubsection{Statistical Characterization for NF Devices}
We assume that $\{\bm h_n\}_{n=1}^N$ are mutually independent across devices and independent of the noise $\bm W$. \revised{Cross-device independence is standard in covariance-based activity detection~\cite{chen2018sparse,fengler2021nonTIT,wangcas2025TWCNearField}.}
While the LoS component is deterministic given the device's location, the NLoS scattering environment is random. We assume the NLoS gains $\{\beta_{n,\ell}\}$, $\ell=1,\ldots,L_n$, are independent zero-mean complex Gaussian variables, i.e., $\beta_{n,\ell}\sim\mathcal{CN}(0,\sigma_{n,\ell}^2),$ where $\sigma_{n,\ell}^2=\mathbb E[|\beta_{n,\ell}|^2]$. %
Since finite linear combinations of independent complex Gaussian variables remain complex Gaussian, the aggregate NLoS component is also complex Gaussian. 
Thus, the total channel vector $\bm{h}_n$ follows a spatially correlated Rician distribution
\begin{equation}
    \bm{h}_n \sim \mathcal{CN}(\bar{\bm{h}}_n, \bm{R}_n), \qquad n\in\mathcal{N}_{\mathrm{NF}}, 
    \label{eq:channel_distribution}
\end{equation}
where the mean is determined by the LoS path:
\begin{equation}
    \bar{\bm{h}}_n \triangleq \mathbb{E}[\bm{h}_n] = \beta_{n,0} \bm{b}(r_n, \theta_n) ,
    \label{eq:mean_def}
\end{equation}
and the spatial covariance matrix is
\begin{align}
    \bm{R}_n & \triangleq \mathbb{E}[(\bm{h}_n - \bar{\bm{h}}_n)(\bm{h}_n - \bar{\bm{h}}_n)^\herm] \notag \\
    & = \sum_{\ell=1}^{L_n} \sigma_{n,\ell}^2 \bm{b}(r_{n,\ell}, \theta_{n,\ell}) \bm{b}(r_{n,\ell}, \theta_{n,\ell})^\herm.
    \label{eq:cov_def}
\end{align}
The covariance matrix in \eqref{eq:cov_def} is a sum of $L_n$ rank-one terms. When the number of dominant scattering clusters is moderate ($L_n \ll M$), the covariance matrix is low-rank and its energy is confined to a subspace determined by the scatterer geometry. %

\subsubsection{FF Devices}
For FF devices ($r_n \gg Z_{\mathrm{Ray}}$), the wavefront curvature is negligible, and the NF steering vector $\bm b(r_n,\theta_n)$ in \eqref{eq:steering_vec} reduces to the classical plane-wave form with entries
\begin{equation}
    [\bm b(r_n,\theta_n)]_m \approx \frac{1}{\sqrt{M}}
    \exp\!\left(-j\frac{2\pi}{\lambda}\delta_m\sin\theta_n\right),
    \label{eq:ff_steering}
\end{equation} %
with $\delta_m = \left( m-\frac{M+1}{2} \right) d$.
\revised{For an FF device, the plane-wave steering vector depends only on the angle, and hence,  the distance argument $r_n$ in $\bm b(r_n,\theta_n)$ is inactive.}
For FF devices, we have
\begin{equation}
    \bm h_n \sim \mathcal{CN}(\bar{\bm h}_n,\bm R_n), \qquad n\in\mathcal{N}_{\mathrm{FF}},
\end{equation}
where the NLoS covariance reduces to a scaled identity
\begin{equation}
    \bar{\bm h}_n = \beta_{n,0}\bm b(r_n,\theta_n), \qquad
    \bm R_n = g_n \bm I_M, \qquad n\in\mathcal{N}_{\mathrm{FF}},
    \label{eq:ff_stats}
\end{equation}
where $g_n>0$ \revised{denotes the per-antenna variance of the residual FF NLoS component.}
\revised{The scaled-identity covariance $\bm R_n=g_n\bm I_M$ is the FF uncorrelated-scattering specialization, where after removing the Rician mean, the residual antenna-domain entries are modeled as uncorrelated with common variance $g_n$. 
This specialization is not used for NF devices, whose covariance is structured, low-rank, and generally non-diagonal.}
The case of $\bar{\bm h}_n = \bm 0$ corresponds to Rayleigh-only FF links.

\revised{The long-term statistics $(\bar{\bm h}_n,\bm R_n)$ are estimated offline. At registration or deployment, each device transmits pilots over several fading blocks, from which the base station estimates the large-scale power, the Rician factor, the LoS angle, and the residual covariance~\cite{Liu2024CWOMMLE}. NF positioning and sensing can provide the LoS angle and distance~\cite{Lu2024NFXLMIMOTutorial,Liu2023NFReview}. These statistics vary slowly and are refreshed only when the environment changes significantly, so their overhead is amortized over many access attempts and is not incurred in the online slots.}

\section{Problem Formulation}
\label{sec:prob_form}

The goal of device activity detection is to recover the binary activity pattern
$\bm{\alpha} = [\alpha_1, \dots, \alpha_N]^\top$ from the received signal matrix $\bm{Y}$,
where $\alpha_n\in\{0,1\}$ indicates whether device $n$ is active.

\subsection{Joint Statistics of the Vectorized Observation}
For the all-FF isotropic case $\bm R_n=g_n\bm I_M$, conditioned on the activity pattern, the mean-removed columns of $\bm Y$ are independent and share the same residual covariance. Covariance-based inference can therefore be carried out through an $L\times L$ sample covariance matrix by treating the $M$ antenna-domain residual observations as independent snapshots.
In mixed NF/FF scenarios with $\mathcal N_{\rm NF}\neq\emptyset$, this simplification is generally no longer valid because the mean-removed columns become statistically dependent across antenna elements.
We define the vectorized received signal $\bm{y} \triangleq \text{vec}(\bm{Y}) \in \mathbb{C}^{LM}$.
Using the identity $\text{vec}(\bm{u}\bm{v}^\top) = \bm{v} \otimes \bm{u}$, the signal model \eqref{eq:received_signal} can be reformulated~as
\begin{align}
    \bm{y} &= \text{vec}\left( \sum_{n=1}^{N} \alpha_n \bm{s}_n \bm{h}_n^\top + \bm{W} \right) \notag \\
    &= \sum_{n=1}^{N} \alpha_n \text{vec}(\bm{s}_n \bm{h}_n^\top) + \text{vec}(\bm{W}) = \sum_{n=1}^{N} \alpha_n (\bm{h}_n \otimes \bm{s}_n) + \bm{w},
    \label{eq:vec_y_derivation}
\end{align}
where $\bm{w} \sim \mathcal{CN}(\bm{0}, \sigma_w^2 \bm{I}_{LM})$.
Conditioned on the channel statistics and the activity pattern $\bm{\alpha}$, $\bm{y}$ follows a multivariate complex Gaussian distribution
\begin{equation}
    \bm{y} \mid \bm{\alpha} \sim \mathcal{CN}(\bm{\mu}_{\bm{\alpha}}, \bm{\Sigma}_{\bm{\alpha}}).
\end{equation}
The aggregate mean vector $\bm{\mu}_{\bm{\alpha}} \in \mathbb{C}^{LM}$ accounts for the deterministic LoS components
\begin{equation} 
    \bm{\mu}_{\bm{\alpha}} = \mathbb{E}[\bm{y} \mid \bm{\alpha}] = \sum_{n=1}^{N} \alpha_n \bm m_n,
    \label{eq:agg_mean}
\end{equation}
where $\bm m_n\triangleq \bar{\bm h}_n\otimes \bm s_n$.
The aggregate covariance matrix $\bm{\Sigma}_{\bm{\alpha}} \in \mathbb{C}^{LM \times LM}$ captures the joint covariance structure induced by the pilots and device channels
\begin{align} 
    \bm{\Sigma}_{\bm{\alpha}} &= \mathbb{E}\left[ (\bm{y} - \bm{\mu}_{\bm{\alpha}})(\bm{y} - \bm{\mu}_{\bm{\alpha}})^\herm \right] \notag \\
    &= \sum_{n=1}^{N} \alpha_n \left( \mathbb{E}[(\bm{h}_n - \bar{\bm{h}}_n)(\bm{h}_n - \bar{\bm{h}}_n)^\herm] \otimes \bm{s}_n \bm{s}_n^\herm \right) + \sigma_w^2 \bm{I}_{LM} \notag \\
    &= \sum_{n=1}^{N} \alpha_n \bm{C}_n + \sigma_w^2 \bm{I}_{LM},
    \label{eq:agg_cov}
\end{align}
where $\bm{C}_n \triangleq \bm{R}_n \otimes \bm{s}_n \bm{s}_n^\herm$ is the per-device covariance matrix in the vectorized observation domain.

\subsection{Relaxed MLE Problem}

Given \eqref{eq:agg_mean} and \eqref{eq:agg_cov}, the conditional likelihood of $\bm y$ under a \emph{binary} activity pattern $\bm\alpha$ is
\begin{equation} \label{eq:likelihood}
p(\bm y \mid \bm\alpha)
= \frac{1}{\pi^{LM}|\bm\Sigma_{\bm\alpha}|}
\exp\left(
-(\bm y-\bm\mu_{\bm\alpha})^\herm \bm\Sigma_{\bm\alpha}^{-1} (\bm y-\bm\mu_{\bm\alpha})
\right),
\end{equation}
where $\bm\mu_{\bm\alpha}=\sum_{n=1}^{N} \alpha_n \bm m_n$ and
$\bm\Sigma_{\bm\alpha}=\sum_{n=1}^{N} \alpha_n \bm C_n+\sigma_w^2 \bm I_{LM}$.
By taking the negative logarithm and removing the constant term $LM\log\pi$, we obtain the negative log-likelihood (NLL)
\begin{equation}
\mathcal{L}(\bm\alpha)
=
\log |\bm{\Sigma}_{\bm{\alpha}}|
+
(\bm{y} - \bm{\mu}_{\bm{\alpha}})^\herm
\bm{\Sigma}_{\bm{\alpha}}^{-1}
(\bm{y} - \bm{\mu}_{\bm{\alpha}}).
\label{eq:nll_alpha}
\end{equation}

The exact MLE-based activity detection amounts to the combinatorial optimization
\begin{equation} \label{eq:binary_ml}
    \begin{split}
        \minimize_{\bm\alpha} \quad & \mathcal{L}(\bm\alpha) \\
        \st \quad & \alpha_n \in\{0,1\}, \quad \forall n = 1, \dots, N,
    \end{split}
\end{equation}
which is intractable for large-scale mMTC with $N\gg 1$.
To obtain a computationally tractable formulation, we relax the binary constraint by introducing $\bm\gamma\in[0,1]^N$ as a continuous counterpart of $\bm\alpha$.
We then define the relaxed NLL by replacing $\bm\alpha$ with $\bm\gamma$ in \eqref{eq:nll_alpha}, namely,
\begin{equation}
\mathcal{L}(\bm\gamma)
=
\log |\bm{\Sigma}_{\bm{\gamma}}|
+
(\bm{y} - \bm{\mu}_{\bm{\gamma}})^\herm
\bm{\Sigma}_{\bm{\gamma}}^{-1}
(\bm{y} - \bm{\mu}_{\bm{\gamma}}),
\label{eq:nll}
\end{equation}
where $\bm\mu_{\bm\gamma}=\sum_{n=1}^{N} \gamma_n \bm m_n$ and
$\bm\Sigma_{\bm\gamma}=\sum_{n=1}^{N} \gamma_n \bm C_n+\sigma_w^2 \bm I_{LM}$.
We focus further on the unregularized relaxed MLE problem
\begin{equation} \label{eq:problem_formulation}
    \begin{split}
        \minimize_{\bm{\gamma}} \quad &
        \mathcal{L}(\bm\gamma)
        \\
        \st \quad & 0 \le \gamma_n \le 1,\quad \forall n = 1, \dots, N.
    \end{split}
\end{equation}
For every feasible $\bm\gamma\in[0,1]^N$, the aggregate covariance satisfies
\begin{equation}
\bm\Sigma_{\bm\gamma}\succeq \sigma_w^2\bm I_{LM},
\end{equation}
so $\bm\Sigma_{\bm\gamma}$ is Hermitian positive definite. Hence $\log|\bm\Sigma_{\bm\gamma}|$ and $\bm\Sigma_{\bm\gamma}^{-1}$ are well defined, and $\mathcal L(\bm\gamma)$ is a smooth but nonconvex objective on the compact box $[0,1]^N$. %
Finally, a binary activity decision can be obtained from the relaxed solution by thresholding, i.e., $\hat\gamma_n^{\sf bin}=\mathbb{I}\{\hat\gamma_n\ge\tau\}$, or by an oracle top-$K$ rule for support-recovery benchmarking when $K$ is known.
Direct minimization of \eqref{eq:problem_formulation} is nevertheless nontrivial because the stationary conditions do not admit a closed-form solution, and each activity variable perturbs both the mean and the covariance. As a result, even first-order methods require repeated evaluations of globally coupled inverse- and log-determinant terms through $\bm\Sigma_{\bm\gamma}$. %

\section{Proposed MM-PGD Framework}
\label{sec:mm_alg}

We develop an MM-PGD method for addressing problem \eqref{eq:problem_formulation}. At each iteration, we construct a quadratic majorizer of the relaxed NLL around the current iterate and minimize it over the box constraint $[0,1]^N$. We first derive the gradient and the MM majorizer, then obtain the projected update with backtracking, and finally describe the exact Kronecker--Woodbury modules used to evaluate the method efficiently.

\subsection{Gradient Structure and Majorization}
\label{subsec:mm_majorize}

Recall the relaxed MLE problem \eqref{eq:problem_formulation}. 
Given $\bm\gamma$, define the aggregate mean and covariance
\begin{equation}
\bm\mu_{\bm\gamma}\triangleq \sum_{n=1}^N \gamma_n\bm m_n,
\qquad
\bm\Sigma_{\bm\gamma}\triangleq \sum_{n=1}^N \gamma_n\bm C_n+\sigma_w^2\bm I_{LM},
\label{eq:mm_defs}
\end{equation}
where $\bm m_n\triangleq \bar{\bm h}_n\otimes \bm s_n$ and $\bm C_n\triangleq \bm R_n\otimes(\bm s_n\bm s_n^\herm)$. We also define
\begin{equation}
\bm r_{\bm\gamma}\triangleq \bm y-\bm\mu_{\bm\gamma},
\qquad
\bm v\triangleq \bm\Sigma_{\bm\gamma}^{-1}\bm r_{\bm\gamma}.
\end{equation}
\begin{proposition}[Gradient of the relaxed NLL]
\label{prop:gradientNLL}
For any feasible $\bm\gamma\in[0,1]^N$, the gradient of \eqref{eq:problem_formulation} has entries
\begin{equation}
\big[\nabla\mathcal L(\bm\gamma)\big]_n
=
\mathrm{tr}\!\big(\bm\Sigma_{\bm\gamma}^{-1}\bm C_n\big)
-
\bm v^\herm \bm C_n \bm v
-
2\,\mathrm{Re}\!\left\{\bm v^\herm \bm m_n\right\},
\label{eq:mm_grad_closed}
\end{equation}
for $n=1,\dots,N$.
\end{proposition}

\begin{proof}
See Appendix~\ref{app:proof_gradientNLL}.
\end{proof}

Moreover, \eqref{eq:mm_grad_closed} shows that the relaxed objective in \eqref{eq:problem_formulation} is globally coupled through the common inverse covariance $\bm\Sigma_{\bm\gamma}^{-1}$ and residual $\bm r_{\bm\gamma}$, so the classical FF rank-one coordinate-update logic no longer applies naturally to \eqref{eq:problem_formulation}. %
The objective in \eqref{eq:problem_formulation} is smooth on $[0,1]^N$, and can be upper-bounded by a quadratic majorizer. 
Indeed, denote the value of the optimization variable $\bm\gamma$ at iteration $t$ as $\bm\gamma^{(t)}$. Then for a standard MM step, we need to construct a majorizer $Q(\bm\gamma;\bm\gamma^{(t)})$ such that
\begin{equation}
    \begin{split}
        & Q(\bm\gamma;\bm\gamma^{(t)})\ \ge\ \mathcal L(\bm\gamma) \quad  \forall\bm\gamma\in[0,1]^N, \\
        & Q(\bm\gamma^{(t)};\bm\gamma^{(t)})=\mathcal L(\bm\gamma^{(t)}).
\end{split}
\label{eq:mm_majorizer_def}
\end{equation}
Assume that $\nabla\mathcal L$ is Lipschitz continuous on $[0,1]^N$. \revised{Let $L_{\mathrm{Lip}}\ge0$ denote a corresponding Lipschitz constant, i.e.,
\[
\|\nabla\mathcal L(\bm\gamma)-\nabla\mathcal L(\tilde{\bm\gamma})\|_2
\le
L_{\mathrm{Lip}}\|\bm\gamma-\tilde{\bm\gamma}\|_2,
\qquad
\forall\bm\gamma,\tilde{\bm\gamma}\in[0,1]^N .
\]
At iteration $t$, $L_t>0$ denotes the quadratic-curvature parameter of the majorizer; equivalently, $1/L_t$ is the projected-gradient step size.} 
The descent lemma for MM then implies that for any $L_t\ge L_{\mathrm{Lip}}$ and any $\bm\gamma\in[0,1]^N$, we have
\begin{equation}
\mathcal L(\bm\gamma)
\le
\underbrace{
\mathcal L(\bm\gamma^{(t)})
+
\nabla \mathcal L(\bm\gamma^{(t)})^{\top}(\bm\gamma-\bm\gamma^{(t)})
+
\frac{L_t}{2}\|\bm\gamma-\bm\gamma^{(t)}\|_2^2
}_{\triangleq\ Q(\bm\gamma;\bm\gamma^{(t)})},
\label{eq:quadratic_majorizer}
\end{equation}
and each MM iteration constructed using $Q(\bm\gamma;\bm\gamma^{(t)})$ instead of $\mathcal L(\bm\gamma)$ does not increase the value of the objective function, thus ensuring monotone descent of the objective sequence.

\subsection{Projected Update and Backtracking}
\label{subsec:mm_projected_update}

At iteration $t$, the MM step minimizes the quadratic majorizer on $[0,1]^N$
\begin{equation}
\bm\gamma^{(t+1)}
=
\argmin_{\bm\gamma \in [0,1]^N}\ 
Q(\bm\gamma;\bm\gamma^{(t)}).
\label{eq:mm_subproblem}
\end{equation}
Substituting the right hand side of \eqref{eq:quadratic_majorizer} into \eqref{eq:mm_subproblem} and solving the problem yields the projected gradient update
\begin{equation}
\bm\gamma^{(t+1)}
=
\Pi_{[0,1]^N}\!\left(
\bm\gamma^{(t)}-\frac{1}{L_t}\nabla\mathcal L(\bm\gamma^{(t)})
\right) . 
\label{eq:mm_pgd_update}
\end{equation}

The Lipschitz constant $L_{\mathrm{Lip}}$ is generally unknown, so backtracking is used. \revised{Within the backtracking loop, $L_t$ denotes the current trial curvature value; once the acceptance test below is met, the same symbol denotes the accepted curvature parameter at iteration $t$.} Given a candidate $L_t$, compute \eqref{eq:mm_pgd_update} and accept the current $L_t$ if
\begin{equation}
\begin{split}
    \mathcal L(\bm\gamma^{(t+1)}) & \le
    \mathcal L(\bm\gamma^{(t)}) + \nabla\mathcal L(\bm\gamma^{(t)})^{\top}(\bm\gamma^{(t+1)}-\bm\gamma^{(t)}) \\
    &\quad + \frac{L_t}{2}\|\bm\gamma^{(t+1)}-\bm\gamma^{(t)}\|_2^2.
\end{split}
\label{eq:mm_accept}
\end{equation}
Otherwise, increase the curvature parameter by a \revised{multiplicative backtracking factor $c_{\mathrm{bt}}>1$}, $L_t\leftarrow c_{\mathrm{bt}} L_t$, and repeat. \revised{The backtracking uses an initial curvature candidate $L_0>0$: at $t=0$ the loop starts from $L_0$, and for $t>0$ it warm-starts from the curvature parameter accepted at the previous iteration.} In the experiments reported in the next section, we use $L_0=20$ and $c_{\mathrm{bt}}=1.5$. \revised{Because the descent lemma guarantees acceptance whenever $L_t\ge L_{\mathrm{Lip}}$, this multiplicative rule keeps the accepted curvature parameters uniformly bounded as $0<L_0\le L_t\le \bar L\triangleq\max\{L_0,c_{\mathrm{bt}}L_{\mathrm{Lip}}\}$.} The following proposition establishes the desired monotone descent and projected stationarity properties for iterates \eqref{eq:mm_pgd_update}.

\begin{proposition}[Monotone descent and projected stationarity of MM-PGD]
\label{prop:mm_convergence}
Assume that $\nabla\mathcal L (\bm\gamma)$ is Lipschitz continuous on $[0,1]^N$ and \revised{that the accepted parameters $L_t$ are generated by the backtracking rule above}. Then the MM-PGD iterates generated by \eqref{eq:mm_pgd_update} satisfy the following properties:
\begin{enumerate}
\item the sequence $\{\mathcal L(\bm\gamma^{(t)})\}_t$ is nonincreasing;
\item the composite objective $F(\bm\gamma)\triangleq \mathcal L(\bm\gamma)+I_{[0,1]^N}(\bm\gamma)$ is bounded below on $[0,1]^N$, and the sequence $\{F(\bm\gamma^{(t)})\}_t$ converges;
\item every accumulation point of $\{\bm\gamma^{(t)}\}_t$ is a first-order projected stationary point of $F (\bm\gamma)$ over $[0,1]^N$.
\item \revised{Let $F_{\inf}\triangleq\inf_{\bm\gamma\in[0,1]^N}F(\bm\gamma)$. For the projected-gradient mapping $G_{L_t}(\bm\gamma^{(t)})\triangleq L_t(\bm\gamma^{(t)}-\bm\gamma^{(t+1)})$ and every iteration horizon $T\ge1$,
\begin{equation}
\min_{0\le t<T}\|G_{L_t}(\bm\gamma^{(t)})\|_2^2
\le
\frac{2\bar L\big(F(\bm\gamma^{(0)})-F_{\inf}\big)}{T}.
\label{eq:mm_stationarity_rate}
\end{equation}}
\end{enumerate}
\end{proposition}

\begin{proof}
See Appendix~\ref{app:mm_convergence}.
\end{proof}

\revised{The bound in \eqref{eq:mm_stationarity_rate} is a worst-case best-iterate stationarity guarantee: it concerns the smallest projected-gradient mapping norm among the first $T$ iterates. In particular, \eqref{eq:mm_stationarity_rate} gives $\min_{0\le t<T}\|G_{L_t}(\bm\gamma^{(t)})\|_2=\mathcal O(1/\sqrt T)$ for the smooth nonconvex problem.}

The complete procedure is summarized in Algorithm~\ref{alg:mm_pgd}.

\begin{algorithm}[!t]
    \caption{MM-PGD for \eqref{eq:problem_formulation}}
    \label{alg:mm_pgd}
    \begin{algorithmic}[1]
        \State \textbf{Input:} $\bm y$, $\{\bm m_n,\bm C_n\}_{n=1}^N$, $\sigma_w^2$; initial $\bm\gamma^{(0)}\in[0,1]^N$; initial $L_0>0$; backtracking factor $c_{\mathrm{bt}}>1$; maximum number of iterations $T$.
        \For{$t=0,1,\dots,T-1$}
            \State Evaluate $\nabla\mathcal L(\bm\gamma^{(t)})$ exactly using the computational modules in Subsection~\ref{subsec:mm_woodbury}.
            \State \revised{Initialize the candidate $L_t$ with $L_0$ for $t=0$ and with the previously accepted value for $t>0$.}
            \Repeat
                \State $
                \bm\gamma^{(t+1)} \leftarrow \Pi_{[0,1]^N}\!\left(
                \bm\gamma^{(t)}-\frac{1}{L_t}\nabla\mathcal L(\bm\gamma^{(t)})
                \right)
                $
                \State Evaluate $\mathcal L(\bm\gamma^{(t+1)})$ exactly using the computational modules in Subsection~\ref{subsec:mm_woodbury}.
                \If{\eqref{eq:mm_accept} is satisfied}
                    \State \textbf{accept} $\bm\gamma^{(t+1)}$
                \Else
                    \State $L_t \leftarrow c_{\mathrm{bt}} L_t$
                \EndIf
            \Until{\eqref{eq:mm_accept} is satisfied}
        \EndFor
        \State \textbf{Output:} final accepted iterate $\hat{\bm\gamma}$ and optional binary decision via thresholding or top-$K$ support selection.
    \end{algorithmic}
\end{algorithm}

\subsection{Exact Kronecker--Woodbury Evaluation}
\label{subsec:mm_woodbury}

The exact evaluation of the gradient in \eqref{eq:mm_grad_closed} is explained below.
For any $\bm x\in\mathbb C^{LM}$, let $\bm X\in\mathbb C^{L\times M}$ satisfy $\bm x=\mathrm{vec}(\bm X)$. Using the property $\mathrm{vec}(\bm A\bm X\bm B)=(\bm B^\top\otimes \bm A)\mathrm{vec}(\bm X)$, we have
\begin{equation}
(\bm R_n\otimes \bm s_n\bm s_n^\herm)\,\mathrm{vec}(\bm X)
=
\mathrm{vec} \left( (\bm s_n\bm s_n^\herm)\bm X\,\bm R_n^\top \right).
\label{eq:mm_kron_apply}
\end{equation}
Consequently,
\begin{equation}
\bm\Sigma_{\bm\gamma}\bm x
=
\sigma_w^2\bm x
+
\sum_{n=1}^N \gamma_n\,(\bm R_n\otimes \bm s_n\bm s_n^\herm)\bm x,
\label{eq:mm_sigma_apply}
\end{equation}
which applies the covariance operator without forming an $LM\times LM$ matrix. Let $\bm v=\mathrm{vec}(\bm V)$ with $\bm V\in\mathbb C^{L\times M}$, and define
\begin{equation}
\bm q_n \triangleq \bm V^\herm \bm s_n\in\mathbb C^{M}.
\label{eq:mm_gn}
\end{equation}
Then, we have
\begin{equation}
\bm v^\herm \bm m_n
=
\bar{\bm h}_n^{\top}\bm q_n,
\label{eq:mm_mean_term}
\end{equation}
and
\begin{equation}
\bm v^\herm \bm C_n \bm v
=
\mathrm{tr}\!\left(\bm V^\herm(\bm s_n\bm s_n^\herm)\bm V\,\bm R_n^\top\right)
=
\bm q_n^\herm \bm R_n^\top \bm q_n.
\label{eq:mm_quad_term}
\end{equation}
If $\bm R_n=\bm U_n\bm\Lambda_n\bm U_n^\herm$ has rank $\tilde r_n\triangleq \operatorname{rank}(\bm R_n)\approx L_n$, then
\begin{equation}
\bm q_n^\herm \bm R_n^\top \bm q_n
=
\left\|\bm\Lambda_n^{1/2}\bm U_n^{\top}\bm q_n\right\|_2^2,
\label{eq:mm_quad_lowrank}
\end{equation}
which lowers the per-user cost of computing the quadratic term.

To turn these per-device expressions into an efficient exact computational module, we next exploit the specific structure of the mixed NF/FF aggregate covariance. Note that FF devices contribute to the covariance matrix a Kronecker-structured term, whereas NF devices contribute a low-rank correction. This structure allows for a Woodbury-compatible decomposition tailored to \eqref{eq:agg_cov}.

\begin{proposition}[Kronecker-plus-low-rank form of the mixed NF/FF covariance]
\label{prop:woodbury_decomp}
Under the mixed NF/FF model with $\bm R_n = g_n \bm I_M$ for $n\in\mathcal N_{\mathrm{FF}}$ and $\bm R_n = \bm U_n\bm\Lambda_n\bm U_n^\herm$ with rank $\tilde r_n$ for $n\in\mathcal N_{\mathrm{NF}}$, the aggregate covariance admits the decomposition
\begin{equation}
\bm\Sigma_{\bm\gamma}
=
\bm I_M \otimes \tilde{\bm A}_{\bm\gamma}
+
\tilde{\bm B}_{\bm\gamma}\tilde{\bm B}_{\bm\gamma}^\herm,
\label{eq:wb_decomp}
\end{equation}
where
\begin{align}
\tilde{\bm A}_{\bm\gamma}
&\triangleq
\sigma_w^2\bm I_L + \sum_{n\in\mathcal N_{\mathrm{FF}}} \gamma_n g_n\, \bm s_n\bm s_n^\herm
\;\in\;\mathbb C^{L\times L},
\label{eq:wb_Atilde}
\\
\tilde{\bm B}_{\bm\gamma}
&\triangleq
\Big[
\sqrt{\gamma_n\lambda_{n,k}}\;(\bm u_{n,k}\otimes \bm s_n)
\Big]
\;\in\;\mathbb C^{LM\times r_{\mathrm{NF}}}, \quad n\in\mathcal N_{\mathrm{NF}}, \nonumber \\
& \qquad \qquad \qquad \qquad \qquad \qquad \qquad \quad k=1,\dots,\tilde r_n,
\label{eq:wb_Btilde}
\end{align}
with total NF rank $r_{\mathrm{NF}}\triangleq\sum_{n\in\mathcal N_{\mathrm{NF}}} \tilde r_n$.
\end{proposition}

\begin{proof} See Appendix~\ref{app:proof_woodbury_decomp}. 
\end{proof}

\revised{The contribution of the exact evaluation is the Kronecker-plus-low-rank structural decomposition of the mixed NF/FF covariance established in Proposition~\ref{prop:woodbury_decomp}, in which the FF devices form the Kronecker term and the NF devices form the low-rank term. This structure makes the matrix determinant lemma and the Woodbury identity applicable, avoiding factorization of the full $LM\times LM$ covariance.}

\noindent \textbf{Exact NLL evaluation via Kronecker--Woodbury.}
Define $\bm W_{\bm\gamma}\triangleq \bm I_{r_{\mathrm{NF}}}+\tilde{\bm B}_{\bm\gamma}^\herm(\bm I_M\otimes \tilde{\bm A}_{\bm\gamma}^{-1})\tilde{\bm B}_{\bm\gamma}\in\mathbb C^{r_{\mathrm{NF}}\times r_{\mathrm{NF}}}$. Using the matrix determinant lemma and the Woodbury identity~\cite{golub2013matrixcomputations}, we have 
\begin{equation}
\log|\bm\Sigma_{\bm\gamma}|
=M\log|\tilde{\bm A}_{\bm\gamma}|
+\log|\bm W_{\bm\gamma}|,
\label{eq:wb_logdet}
\end{equation}
where each factor is obtained by Cholesky decomposition of $\tilde{\bm A}_{\bm\gamma}\in\mathbb C^{L\times L}$ and $\bm W_{\bm\gamma}\in\mathbb C^{r_{\mathrm{NF}}\times r_{\mathrm{NF}}}$.

For the whitened residual, we have
\begin{equation}
\bm v
=
\bm\Sigma_{\bm\gamma}^{-1}\bm r_{\bm\gamma}
=
\bar{\bm v}
-(\bm I_M\!\otimes\! \tilde{\bm A}_{\bm\gamma}^{-1})\tilde{\bm B}_{\bm\gamma}\,\bm W_{\bm\gamma}^{-1}\,\tilde{\bm B}_{\bm\gamma}^\herm\bar{\bm v},
\label{eq:wb_v}
\end{equation}
where $\bar{\bm v}\triangleq (\bm I_M\otimes \tilde{\bm A}_{\bm\gamma}^{-1})\bm r_{\bm\gamma}$. Using Cholesky decomposition of $\tilde{\bm A}_{\bm\gamma}$, the Kronecker product decomposes into $M$ independent $L\times L$ triangular matrices. 

For the trace terms in \eqref{eq:mm_grad_closed}, for FF devices, we have
\begin{equation}
\tr(\bm\Sigma_{\bm\gamma}^{-1}\bm C_n)=
g_n\!\left(
M\,\bm s_n^\herm \tilde{\bm A}_{\bm\gamma}^{-1}\bm s_n
-\tr\big(\bm Q_n^\herm\bm W_{\bm\gamma}^{-1}\bm Q_n\big)
\right),
\label{eq:wb_trace_ff}
\end{equation}
where
\begin{equation}
\bm Q_n \triangleq \tilde{\bm B}_{\bm\gamma}^\herm(\bm I_M\otimes\tilde{\bm A}_{\bm\gamma}^{-1})(\bm I_M\otimes\bm s_n)\in\mathbb C^{r_{\mathrm{NF}}\times M}.
\end{equation}
For NF devices, we have
\begin{equation}
\tr(\bm\Sigma_{\bm\gamma}^{-1}\bm C_n)
=
\tr(\bm R_n)\,\bm s_n^\herm \tilde{\bm A}_{\bm\gamma}^{-1}\bm s_n
-\sum_{k=1}^{\tilde r_n}\lambda_{n,k}\,\bm p_{n,k}^\herm\bm W_{\bm\gamma}^{-1}\bm p_{n,k},
\label{eq:wb_trace_nf}
\end{equation}
where
\begin{equation}
\bm b_{n,k}\triangleq \bm u_{n,k}\otimes\bm s_n,\quad
\bm p_{n,k}\triangleq\tilde{\bm B}_{\bm\gamma}^\herm(\bm I_M\otimes\tilde{\bm A}_{\bm\gamma}^{-1})\bm b_{n,k}\in\mathbb C^{r_{\mathrm{NF}}}.
\end{equation}
The backtracking condition \eqref{eq:mm_accept} therefore evaluates the exact NLL~\eqref{eq:nll} without ever factorizing the full $LM\times LM$ covariance matrix.

For running the backtracking, the trace terms \eqref{eq:wb_trace_ff} and \eqref{eq:wb_trace_nf} are not needed because only $\mathcal L(\bm\gamma)$ must be reevaluated. The Woodbury factors and $\bm v$ computed for accepted $L_t$ are therefore cached and reused when the next gradient is evaluated. This caching improves runtime, while it does not change the exact objective evaluation or the backtracking condition \eqref{eq:mm_accept}.

\noindent \textbf{Offline statistics and accepted-state recursion.}
The per-device statistics $\{\bm m_n,\bm C_n\}_{n=1}^N$ are fixed once $(\bar{\bm h}_n,\bm R_n,\bm s_n)$ are specified. 
Moreover, we can precompute the eigendecompositions of $\bm R_n=\bm U_n\bm\Lambda_n\bm U_n^\herm$, $\forall n$ and the FF/NF partition $\mathcal N_{\mathrm{FF}}$ and $\mathcal N_{\mathrm{NF}}$.

Assuming uniform activity-rate prior, $\bm\gamma$ is initialized as $\bm\gamma^{(0)}=(K/N)\,\bm 1_N$. The initial $\bm\mu$, $\bm\Sigma$, and $\bm r$ can then be computed, respectively, as: 
\begin{equation}
\begin{split}
    \bm\mu^{(0)} & =\sum_{n=1}^N \gamma_n^{(0)}\bm m_n, \\
    \bm\Sigma^{(0)} & =\sigma_w^2\bm I_{LM}+\sum_{n=1}^N \gamma_n^{(0)}\bm C_n, \\
    \bm r^{(0)} & =\bm y-\bm\mu^{(0)}.
\end{split}
\end{equation}
The corresponding $\bm v^{(0)}$ is obtained from $\bm\Sigma^{(0)}\bm v^{(0)}=\bm r^{(0)}$. In subsequent iterations, the mean vector $\bm\mu$ and the covariance matrix $\bm\Sigma$ can be iteratively updated as: 
\begin{equation}
\begin{split}
\bm\mu^{(t+1)} & =
\bm\mu^{(t)}+\sum_{n=1}^N \Delta\gamma_n^{(t)}\,\bm m_n, \\
\bm\Sigma^{(t+1)} & =
\bm\Sigma^{(t)}+\sum_{n=1}^N \Delta\gamma_n^{(t)}\,\bm C_n,
\end{split}
\label{eq:mm_state_update}
\end{equation}
where $\Delta\bm\gamma^{(t)}\triangleq \bm\gamma^{(t+1)}-\bm\gamma^{(t)}$. The residual $\bm r$ and $\bm v$ are then updated as: 
\begin{equation}
\begin{split}
\bm r^{(t+1)} &=\bm y-\bm\mu^{(t+1)}, \\
\bm v^{(t+1)} &\ \text{from solving}\ \bm\Sigma^{(t+1)}\bm v^{(t+1)}=\bm r^{(t+1)}.
\end{split}
\label{eq:mm_v_update}
\end{equation}
Therefore, $\bm\Sigma_{\bm\gamma}$ does not need to be formed explicitly, and the main computations are carried out through iterative updates. This reduces the computational cost. %

\subsection{Complexity Comparison and Special Cases}
\label{subsec:mm_complexity}

One MM-PGD iteration with the exact Kronecker--Woodbury evaluation consists of the following three stages.
\begin{enumerate}
\item \textbf{Core factor construction.} Forming $\tilde{\bm A}_{\bm\gamma}$ requires $\mathcal O(N_{\mathrm{FF}}L^2)$ operations, followed by an $L\times L$ Cholesky decomposition at cost $\mathcal O(L^3)$. Assembling $\tilde{\bm B}_{\bm\gamma}$ in \eqref{eq:wb_Btilde} and constructing $\bm W_{\bm\gamma}$ has the cost $\mathcal O(LM\,r_{\mathrm{NF}}^2)$, while $\bm W_{\bm\gamma}$ is factorized at cost $\mathcal O(r_{\mathrm{NF}}^3)$.
\item \textbf{Exact gradient and likelihood evaluation.} Using the factorizations, $\log|\bm\Sigma_{\bm\gamma}|$ is evaluated via \eqref{eq:wb_logdet}, $\bm v$ is computed via \eqref{eq:wb_v}, the trace terms are evaluated via \eqref{eq:wb_trace_ff} and \eqref{eq:wb_trace_nf}, and $\nabla\mathcal L(\bm\gamma)$ is finally computed via \eqref{eq:mm_grad_closed}. The remaining per-user computations scale linearly with $N$ and are lower-order than the matrix-factorization cost under the tested settings.
\item \textbf{Projected update and backtracking.} Element-wise clipping onto $[0,1]$ in \eqref{eq:mm_pgd_update} has negligible cost. For each backtracking iteration, the computations reduce to rebuilding the candidate-specific Woodbury factors and reevaluating the exact NLL, while the trace computations are needed only for evaluating the gradient.
\end{enumerate}
The dominant matrix-factorization cost per MM-PGD iteration is therefore
\begin{equation}
\mathcal O(L^3+r_{\mathrm{NF}}^3+LM\,r_{\mathrm{NF}}^2),
\end{equation}
which can be much lower than the $\mathcal O((LM)^3)$ cost of factorizing the full $LM\times LM$ covariance when $r_{\mathrm{NF}}\ll LM$.

The relevant comparisons are with~\cite{Liu2024CWOMMLE, wangcas2025TWCNearField}. In~\cite{Liu2024CWOMMLE}, the isotropic FF model requires the computations only in $L\times L$ covariance domain for evaluating the likelihood. Updating one activity variable then induces a rank-one covariance perturbation, so Sherman--Morrison recursion is applicable, and no $LM\times LM$ factorization is needed. In \revised{NF-CD}~\cite{wangcas2025TWCNearField}, the likelihood is written for the vectorized $LM$-dimensional observation and each device contributes a rank-$\tilde r_n$ covariance perturbation, where $\tilde r_n=\operatorname{rank}(\bm R_n)$ is the structural rank of the $n$th device covariance matrix. Sherman--Morrison recursion is no longer applicable, so each coordinate update becomes a higher-rank subproblem with reported inverse-update complexity $\mathcal O((LM)^2 \tilde r_n+\tilde r_n^3)$ until repeated coordinate visits over a full sweep are counted.

By contrast, MM-PGD updates the full activity vector jointly and reduces the exact computation to evaluating two smaller factors $\tilde{\bm A}_{\bm\gamma}\in\mathbb C^{L\times L}$ and $\bm W_{\bm\gamma}\in\mathbb C^{r_{\mathrm{NF}}\times r_{\mathrm{NF}}}$, which is computationally more favorable than the corresponding exact evaluation in \revised{NF-CD}~\cite{wangcas2025TWCNearField}. %

The proposed framework and the corresponding algorithm cover all special cases. In the all-FF case, there is no need for low-rank correction, and the exact evaluation reduces to a single $L\times L$ Cholesky factorization with complexity $\mathcal O(L^3)$. In the mixed NF/FF case, the FF devices contribute to $\tilde{\bm A}_{\bm\gamma}$ while the NF devices contribute to $\tilde{\bm B}_{\bm\gamma}$, yielding the hybrid exact-evaluation complexity $\mathcal O(L^3+r_{\mathrm{NF}}^3+LM\,r_{\mathrm{NF}}^2)$. In the all-NF case, $r_{\mathrm{NF}}$ may approach $LM$, in which case the dominant cost reverts toward full-dimensional factorization. 
\revised{The total NF rank $r_{\mathrm{NF}}$ is not independent of the registered pool size $N$: at a fixed NF fraction $\eta_{\mathrm{NF}}$ and average scatterer count $\bar L_n$, the number of NF devices grows linearly with $N$, so $r_{\mathrm{NF}}\approx\eta_{\mathrm{NF}}\bar L_n N$ and the per-iteration cost grows accordingly. The Kronecker--Woodbury advantage is therefore most pronounced when $r_{\mathrm{NF}}\ll LM$ and shrinks as the pool grows. For example, with $L=30$, $M=64$, $\bar L_n=8$, and $\eta_{\mathrm{NF}}=0.5$, the estimate reaches $LM=1920$ near $N\approx500$.}
The same implementation therefore bridges the classical FF covariance-domain evaluation and the high-rank NF regime without explicit solver-mode switching or regime-specific redesign.

\section{Numerical Experiments}
\label{sec:exp_setup}

This section describes the simulation setup, evaluation protocol, and numerical results.
Unless stated otherwise, all results use the default parameter set in Table~\ref{tab:default_params} and are averaged over independent Monte Carlo  (MC) realizations.

\subsection{Experimental Setup}

For each operating point (e.g., a specific $\mathrm{SNR}_{\mathrm{dB}}$ or pilot length $L$), we run $N_{\mathrm{MC}}=1000$ independent MC trials.
In each trial, the device geometry, NF/FF partition, pilots, activity pattern, channels, and noise are redrawn once and then shared by all compared methods.
Unless a variable is explicitly swept, the default operating point is $(N,K,M,L)=(200,25,64,30)$ with $\kappa_{\mathrm{dB}}=-5$, $\eta_{\mathrm{NF}}=0.5$, and $N_{\mathrm{MC}}=1000$.

\begin{table}[!t]
\centering
\caption{Default simulation parameters.}
\label{tab:default_params}
\renewcommand{\arraystretch}{1.2}
\begin{tabularx}{\linewidth}{l|c|c|X}
\hline
\hline
\textbf{Parameter} & \textbf{Symbol} & \textbf{Default} & \textbf{Description} \\
\hline
\hline
\# devices & $N$ & $200$ & Total potential devices \\
\# active devices & $K$ & $25$ & Active set size (12.5\%) \\
\# BS antennas & $M$ & $64$ & ULA with spacing $d=\lambda/2$ \\
Pilot length & $L$ & $30$ & Pilot dimension \\
Carrier frequency & $f_c$ & $3$ GHz & $\lambda=c/f_c$ \\
Cell radius & $R$ & $500$ m & Max device range \\
NF-user fraction & $\eta_{\mathrm{NF}}$ & $0.5$ & Fraction of NF devices \\
NF placement & -- & annulus & Distance-consistent partition \\
\# scatter clusters & $L_n$ & $8$ & Per-device NLoS clusters \\
Rician $K$-factor & $\kappa_{\mathrm{dB}}$ & $-5$ dB & LoS strength \\
MC trials & $N_{\mathrm{MC}}$ & $1000$ & Unless otherwise stated\\
\hline
\hline
\end{tabularx}
\end{table}

\subsubsection{Pilot Design and Activity Generation}
\label{subsec:exp_pilots}

Each device $n$ is assigned a length-$L$ pilot $\bm s_n\in\mathbb C^L$ with $\|\bm s_n\|_2=1$.
We generate quadrature phase-shift keying (QPSK) pilots with entries
\begin{equation}
[\bm s_n]_\ell \in \left\{\pm \frac{1}{\sqrt{2L}} \pm j\frac{1}{\sqrt{2L}}\right\},\qquad \ell=1,\dots,L.
\end{equation}
\revised{QPSK pilots are constant-modulus, which is favorable for the power amplifiers of low-cost devices. In the likelihood, each pilot enters the mean term through $\bm m_n=\bar{\bm h}_n\otimes\bm s_n$ and the covariance term through $\bm C_n=\bm R_n\otimes\bm s_n\bm s_n^\herm$. We therefore adopt QPSK as a practical finite-alphabet choice, consistent with prior covariance-based activity-detection studies using QPSK, Gaussian, or Bernoulli pilots~\cite{wangcas2025TWCNearField,chen2018sparse,fengler2021nonTIT,marata2024activity}.}
For each realization, an active set $\mathcal K$ of size $K$ is sampled uniformly without replacement, and
$\alpha_n=\mathbb I\{n\in\mathcal K\}$.

\subsubsection{Geometry and Near-/Far-Field Device Composition}
\label{subsec:exp_geometry}

In each Monte Carlo trial, every device is assigned a polar location $(r_n,\theta_n)$ in the BS coordinate system of Section~\ref{sec:sysmd}.
We prescribe the NF-user fraction $\eta_{\mathrm{NF}}\in[0,1]$ and explicitly state $(M,R,\eta_{\mathrm{NF}})$ in each experiment.

\paragraph{Controlled NF/FF composition}
For a given $M$ and $R>Z_{\mathrm{Ray}}(M)$, we draw $\eta_{\mathrm{NF}}N$ devices uniformly in area from the RNF annulus
$r\in(r_{\mathrm{RNF}}(M),Z_{\mathrm{Ray}}(M))$ and draw the remaining $(1-\eta_{\mathrm{NF}})N$ devices uniformly in area from the FF annulus
$r\in[Z_{\mathrm{Ray}}(M),R]$ (all with i.i.d.\ $\theta\sim\mathrm{Unif}[0,\pi]$).
Setting $\eta_{\mathrm{NF}}=1$ (resp.\ $\eta_{\mathrm{NF}}=0$) recovers the all-NF (resp.\ all-FF) special case.
Unless otherwise stated, all reported results use this annulus-based construction to preserve the geometry-consistent NF/FF partition.

\paragraph{Channel generation and normalization}
Within each trial, scatterers are placed uniformly at random in a disk centered at the BS with radius $R_{\mathrm{sc}}$ (default: $200$~m).
For NF devices, the NLoS covariance $\bm R_n$ is constructed as a weighted sum of rank-one outer products of the BS array responses associated with the scatterer-to-BS links, where the weights are given by the product of the two-hop path-loss gains (device$\to$scatterer and scatterer$\to$BS). The channel gains follow the path-loss model $128.1+37.6\log_{10}(d)$~\cite{wangcas2025TWCNearField}, where $d$ is the corresponding distance in km. \revised{We define the per-device average channel power by $\mathbb E\|\bm h_n\|_2^2=\rho_n M$.} Unless otherwise stated, power control sets $\rho_n\equiv\rho_0$ for all devices. For FF devices, we use the isotropic specialization $\bm R_n=g_n\bm I_M$ as in Section~\ref{sec:sysmd}.

Given $(r_n,\theta_n)$, we form the near-field steering vector $\bm b(r_n,\theta_n)$ using \eqref{eq:steering_vec}--\eqref{eq:dist_exact}. We then generate channels $\bm h_n\sim\mathcal{CN}(\bar{\bm h}_n,\bm R_n)$ as
\begin{equation}
\bm h_n = \bar{\bm h}_n + \bm R_n^{1/2}\bm z_n,\qquad \bm z_n\sim\mathcal{CN}(\bm 0,\bm I_M).
\end{equation}
We use $\kappa$ (linear) to denote the Rician $K$-factor and $\kappa_{\mathrm{dB}}\triangleq 10\log_{10}\kappa$ in dB:
\begin{equation}
\kappa \triangleq \frac{\|\bar{\bm h}_n\|_2^2}{\mathrm{tr}(\bm R_n)}.
\label{eq:kappa_def}
\end{equation}
With this definition, we normalize the mean/covariance split via
\begin{equation}
\|\bar{\bm h}_n\|_2^2 = \frac{\kappa}{\kappa+1}\rho_n M,
\qquad
\mathrm{tr}(\bm R_n)=\frac{1}{\kappa+1}\rho_n M.
\label{eq:kappa_split}
\end{equation}
The Rayleigh case corresponds to $\kappa=0$ (equivalently $\kappa_{\mathrm{dB}}=-\infty$), implemented by $\bar{\bm h}_n=\bm 0$ and $\mathrm{tr}(\bm R_n)=\rho_n M$.
\revised{For the FF isotropic specialization $\bm R_n=g_n\bm I_M$, this normalization gives $g_n=\rho_n/(\kappa+1)$; in the Rayleigh-only case $\kappa=0$, it reduces to $g_n=\rho_n$.}

Using $\|\bm b(r_n,\theta_n)\|_2^2=1$, we set
\begin{equation}
\bar{\bm h}_n = \sqrt{\frac{\kappa}{\kappa+1}\rho_n M}\,e^{j\phi_n}\,\bm b(r_n,\theta_n),
\quad
\phi_n\sim\mathrm{Unif}[0,2\pi).
\end{equation}

We report a receive signal-to-noise ratio (SNR) defined as the average per-entry SNR
\begin{equation}
\mathrm{SNR}_{\mathrm{dB}}
\triangleq
10\log_{10}\frac{\bigl\|\sum_{n=1}^N \alpha_n\,\bm s_n\bm h_n^\top\bigr\|_F^2}{L M\,\sigma_w^2}.
\label{eq:snr_def}
\end{equation}
In simulations, for each realization we set $\sigma_w^2$ so that \eqref{eq:snr_def} equals the desired operating point, and then draw $\bm W$ accordingly.

\subsubsection{Evaluation Metrics and Operating Point Selection}
\label{subsec:exp_metrics}

We evaluate activity detection as a support recovery task.
Let $\mathcal K\subset\{1,\dots,N\}$ denote the true active set with $|\mathcal K|=K$, and let $\hat{\bm\gamma}\in\mathbb R^N$ be the soft output (activity score) of an algorithm.
For oracle top-$K$ support-recovery benchmarking, when $K$ is assumed to be known, we declare the detected active set $\hat{\mathcal K}$ as the indices of the $K$ largest entries of $\hat{\bm\gamma}$.
We then compute the miss-detection probability $P_{\mathrm{MD}}$ and false-alarm probability $P_{\mathrm{FA}}$ as
\begin{equation}
P_{\mathrm{MD}} = 1-\frac{|\hat{\mathcal K}\cap \mathcal K|}{K},
\qquad
P_{\mathrm{FA}} = \frac{|\hat{\mathcal K}\setminus \mathcal K|}{N-K}.
\end{equation}
Since $|\hat{\mathcal K}|=|\mathcal K|=K$ under the top-$K$ rule, the number of false positives ($\#\mathrm{FP}$) equals the number of misses ($\#\mathrm{Miss}$)
\begin{equation}
\#\mathrm{FP}=\#\mathrm{Miss}
\quad\Rightarrow\quad
P_{\mathrm{FA}}(N-K)=P_{\mathrm{MD}}K.
\end{equation}
Unless otherwise stated, all main-text plots use this oracle top-$K$ benchmarking protocol.
Accordingly, the reported comparisons focus on support-recovery quality under a common oracle-$K$ assumption.
All $P_{\mathrm{MD}}$ plots use a logarithmic vertical axis; data points where $P_{\mathrm{MD}}=0$ (perfect detection) cannot be displayed on this scale and are omitted.

We compare our proposed MM-PGD (Algorithm~\ref{alg:mm_pgd} with the exact Kronecker--Woodbury implementation in Section~\ref{subsec:mm_woodbury}) with the following methods:
\begin{itemize}
\item CWO-MMLE~\cite{Liu2024CWOMMLE}: a mismatched MLE baseline that replaces $\bm R_n$ by an isotropic approximation $g_n\bm I_M$ with $g_n\triangleq \tr(\bm R_n)/M$ (while retaining $\bar{\bm h}_n$), implemented via coordinate-wise updates.
\item Coordinatewise optimization algorithm (CWO)~\cite[Algorithm 1]{fengler2021nonTIT}
\item Covariance-based matching pursuit (CL-MP)~\cite{marata2024activity}
\item Sparse Bayesian learning (SBL)~\cite{wipf2004sparse}
\item Simultaneous normalized iterative hard thresholding (SNIHT)~\cite[Algorithm 1]{blanchard2014greedy}
\item Simultaneous orthogonal matching pursuit (SOMP)~\cite[Algorithm 3.1]{tropp2006algorithms}
\end{itemize}
These baselines cover coordinate-wise Rician-MLE, covariance-only sparse-recovery, and greedy MMV detectors.
\revised{Among the baselines, SBL, SNIHT, and SOMP use only the pilot observation, CWO and CL-MP use a covariance model without the near-field structure or the LoS mean, and CWO-MMLE retains the LoS mean but replaces $\bm R_n$ by $g_n\bm I_M$.}
In wall-clock terms, the direct full Cholesky implementation of MM-PGD is the most expensive solver in the large-$LM$ regimes studied here, which motivates the exact Kronecker--Woodbury acceleration validated in Section~\ref{app:wb_benchmark}. 
The code is available at \url{https://github.com/xnnjw/mm-pgd-mixed-nf-ff}.

\subsection{Numerical Results}
\label{subsec:results}

We report eight experiments organized into three groups. Experiments~1--4 establish baseline detection behavior under variations in SNR, LoS strength, pilot budget, and activity rate. Experiments~5--7 probe the structural sensitivity of the mixed NF/FF model through array size, NF-user fraction, and scatterer richness. Experiment~8 examines scalability with the total registered device pool. For Experiments~2--6 and Experiment~8, our results are shown at two SNR operating points ($\mathrm{SNR}_{\mathrm{dB}}\in\{0,5\}$) to demonstrate both noise-limited and moderate-SNR performance. Detection performance is reported in this section, whereas the wall-clock validation of the exact Kronecker--Woodbury realization is deferred to Section~\ref{app:wb_benchmark}. In all plots, MM-PGD is shown in red and the vertical axis reports $P_{\mathrm{MD}}$ on a logarithmic scale.

\subsubsection{Experiment~1: SNR sweep}

\begin{figure*}[!t]
    \centering
    \subfloat[$\eta_{\mathrm{NF}}=0.25$\label{fig:exp1_nf25}]{%
        \includegraphics[width=0.24\textwidth]{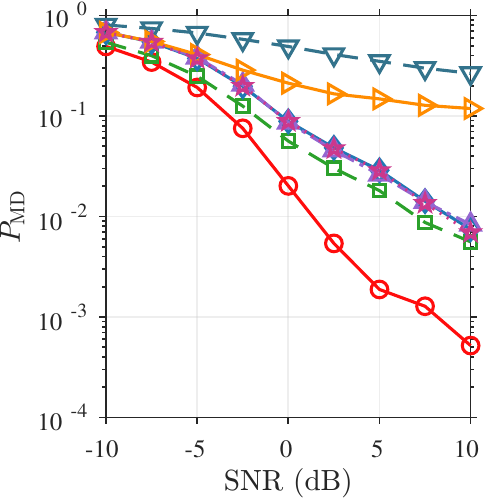}}
    \hfill
    \subfloat[$\eta_{\mathrm{NF}}=0.50$\label{fig:exp1_nf50}]{%
        \includegraphics[width=0.24\textwidth]{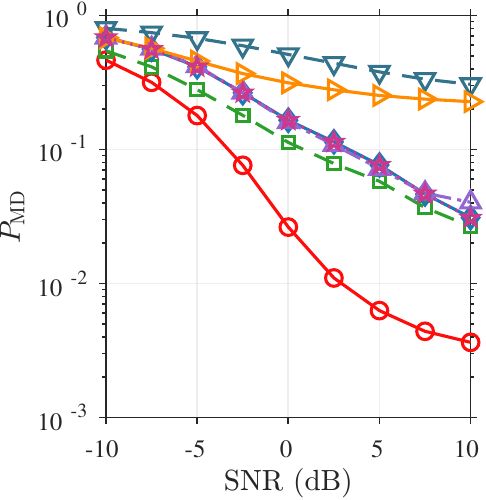}}
    \hfill
    \subfloat[$\eta_{\mathrm{NF}}=0.75$\label{fig:exp1_nf75}]{%
        \includegraphics[width=0.24\textwidth]{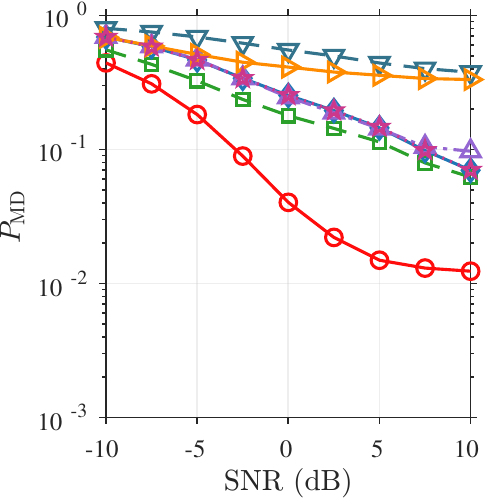}}
    \hfill
    \subfloat[$\eta_{\mathrm{NF}}=1.00$\label{fig:exp1_nf100}]{%
        \includegraphics[width=0.24\textwidth]{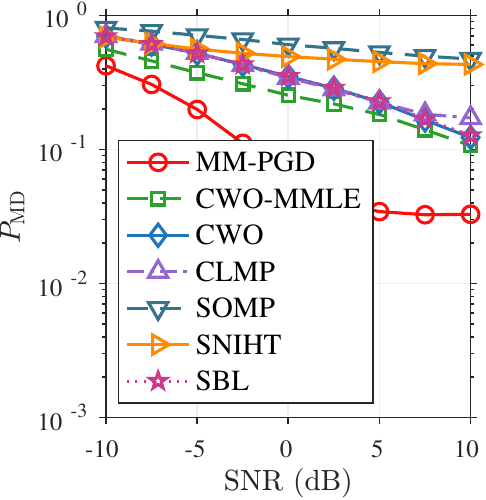}}
    \caption{$P_{\mathrm{MD}}$ vs.\ $\mathrm{SNR}_{\mathrm{dB}}$ (Experiment~1) for four values of the NF-user fraction.
    $(N,K,M,L)=(200,25,64,30)$, $\kappa_{\mathrm{dB}}=-10$, $N_{\mathrm{MC}}=1000$. The proposed MM-PGD gives the lowest $P_{\mathrm{MD}}$. Moreover, the performance gap becomes more visible as the NF-user fraction grows.}
    \label{fig:exp1}
\end{figure*}

Fig.~\ref{fig:exp1} sweeps the SNR from $-10$ to $10$\,dB under a weak-LoS regime ($\kappa_{\mathrm{dB}}=-10$) for four values of the NF-user fraction $\eta_{\mathrm{NF}}\in\{0.25,0.50,0.75,1.00\}$.
Across all four settings, MM-PGD gives the lowest $P_{\mathrm{MD}}$ among the compared methods. %
The main trend is not only that larger SNR helps, but also that the performance gap becomes more visible as the NF-user fraction grows: the all-NF case is harder for every detector, yet MM-PGD retains a clear margin over the strongest coordinate-wise baseline.
This behavior is consistent with the view that, once NF covariance becomes a larger part of the observation structure, explicit mixed-field covariance modeling and full-vector updates become more beneficial.

\subsubsection{Experiment~2: Rician K-factor sweep}

\begin{figure}[!t]
    \centering
    \includegraphics[width=\linewidth]{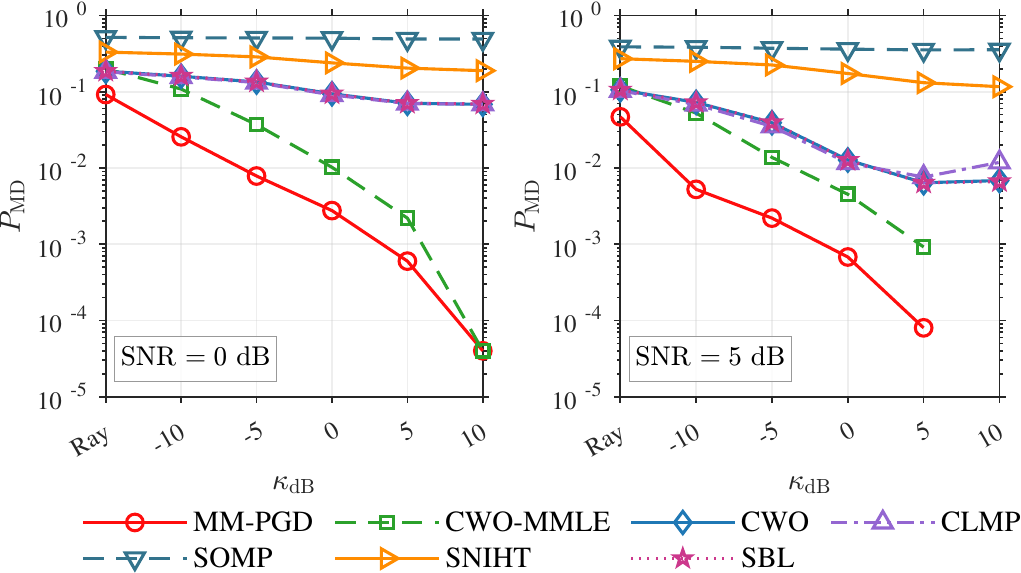}
    \caption{$P_{\mathrm{MD}}$ vs.\ Rician $K$-factor $\kappa_{\mathrm{dB}}$ (Experiment~2).
    Left: $\mathrm{SNR}=0$\,dB; Right: $\mathrm{SNR}=5$\,dB.
    $(N,K,M,L)=(200,25,64,30)$, $\eta_{\mathrm{NF}}=0.5$, $N_{\mathrm{MC}}=1000$. When LoS is not dominant, MM-PGD maintains the largest performance improvement over other methods, whereas the gap naturally shrinks once the LoS becomes dominant.}
    \label{fig:exp2}
\end{figure}

Fig.~\ref{fig:exp2} sweeps the Rician $K$-factor from pure Rayleigh (i.e., zero LoS component) to strong LoS ($\kappa_{\mathrm{dB}}=10$).
All methods that explicitly exploit the LoS mean improve as $\kappa_{\mathrm{dB}}$ increases, with MM-PGD remaining best across the entire range.
The weak-LoS regime from Rayleigh to mildly Rician fading is particularly relevant here. In this regime, the LoS mean is not dominant, and the mixed NF/FF covariance still matters for detection. %
In that regime, MM-PGD maintains the largest gap, whereas the gap naturally shrinks once the LoS component becomes so strong that detection is easy for every mean-aware method.

\subsubsection{Experiment~3: Pilot-length sweep}

\begin{figure}[!t]
    \centering
    \includegraphics[width=\linewidth]{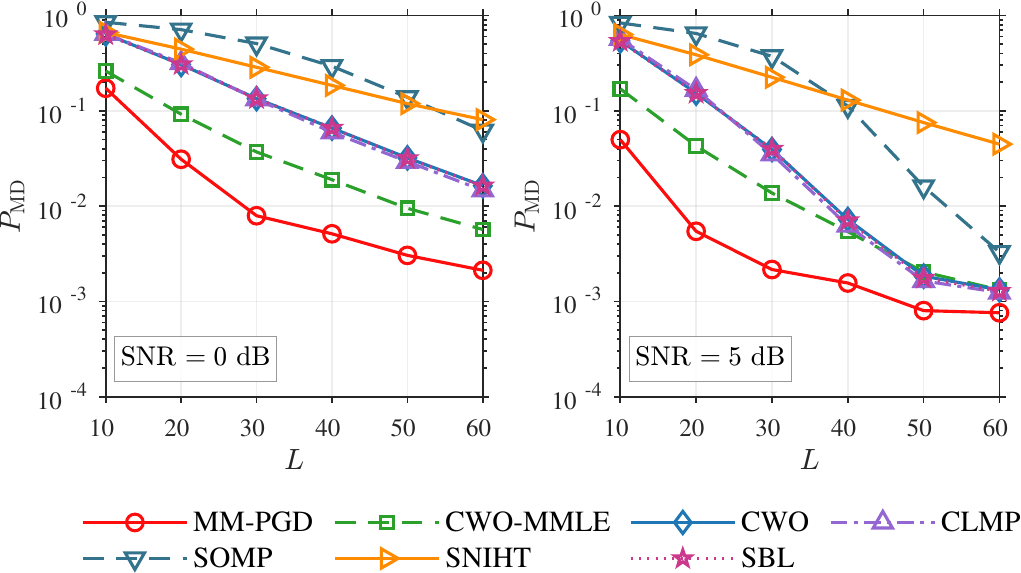}
    \caption{$P_{\mathrm{MD}}$ vs.\ pilot length $L$ (Experiment~3).
    Left: $\mathrm{SNR}=0$\,dB; Right: $\mathrm{SNR}=5$\,dB.
    $(N,K,M)=(200,25,64)$, $\kappa_{\mathrm{dB}}=-5$, $\eta_{\mathrm{NF}}=0.5$, $N_{\mathrm{MC}}=1000$. All methods improve with longer pilots, but MM-PGD has low $P_{\mathrm{MD}}$ with shorter pilots.}
    \label{fig:exp3}
\end{figure}

Fig.~\ref{fig:exp3} sweeps the pilot length from $L=10$ ($N/L=20$) to $L=60$ ($N/L\approx 3.3$).
All methods improve with longer pilots, but MM-PGD reaches the low-$P_{\mathrm{MD}}$ regime with shorter pilot lengths.
The advantage is therefore most visible when the pilot budget is limited.
As $L$ increases, the curves move closer and all methods eventually enter the low-$P_{\mathrm{MD}}$ regime.

\subsubsection{Experiment~4: Activity-rate sweep}

\begin{figure}[!t]
    \centering
    \includegraphics[width=\linewidth]{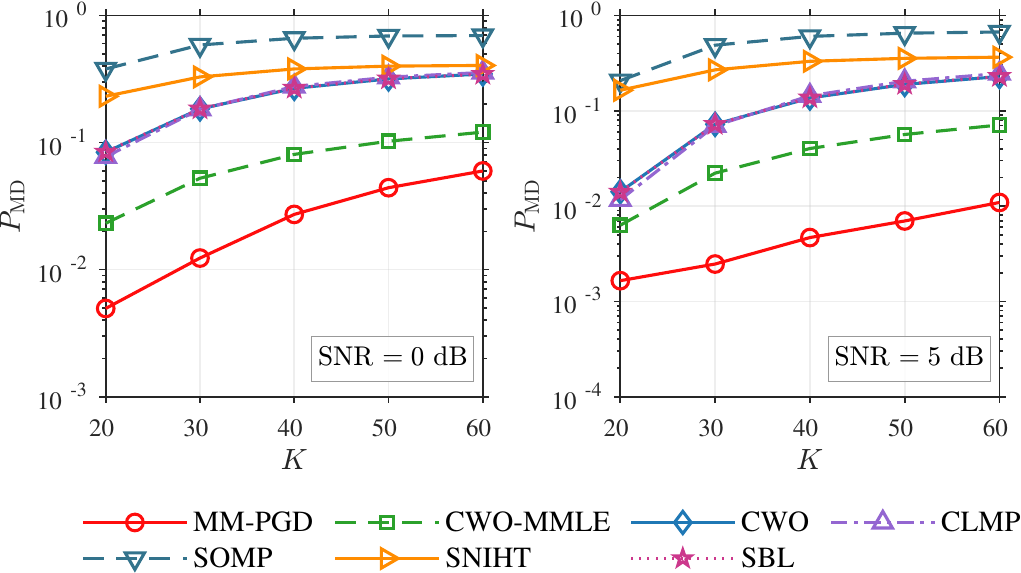}
    \caption{$P_{\mathrm{MD}}$ vs.\ number of active devices $K$ (Experiment~4).
    Left: $\mathrm{SNR}=0$\,dB; Right: $\mathrm{SNR}=5$\,dB.
    $(N,M,L)=(200,64,30)$, $\kappa_{\mathrm{dB}}=-5$, $\eta_{\mathrm{NF}}=0.5$, $N_{\mathrm{MC}}=1000$. All methods degrade as the activity rate increases, but MM-PGD degrades the most gracefully.}
    \label{fig:exp4}
\end{figure}

Fig.~\ref{fig:exp4} varies the number of active devices from $K=10$ (5\% activity) to $K=60$ (30\%).
All methods degrade as the activity rate increases, but MM-PGD degrades the most gracefully.
This trend is consistent with the role of the unified mixed-field likelihood: as more users become simultaneously active, inter-user coupling becomes stronger and the cost of covariance mismatch becomes more visible.
Accordingly, the advantage of MM-PGD over both mismatched Rician-MLE and covariance-only baselines persists throughout the entire activity range.

\subsubsection{Experiment~5: Antenna-count sweep}

\begin{figure}[!t]
    \centering
    \includegraphics[width=\linewidth]{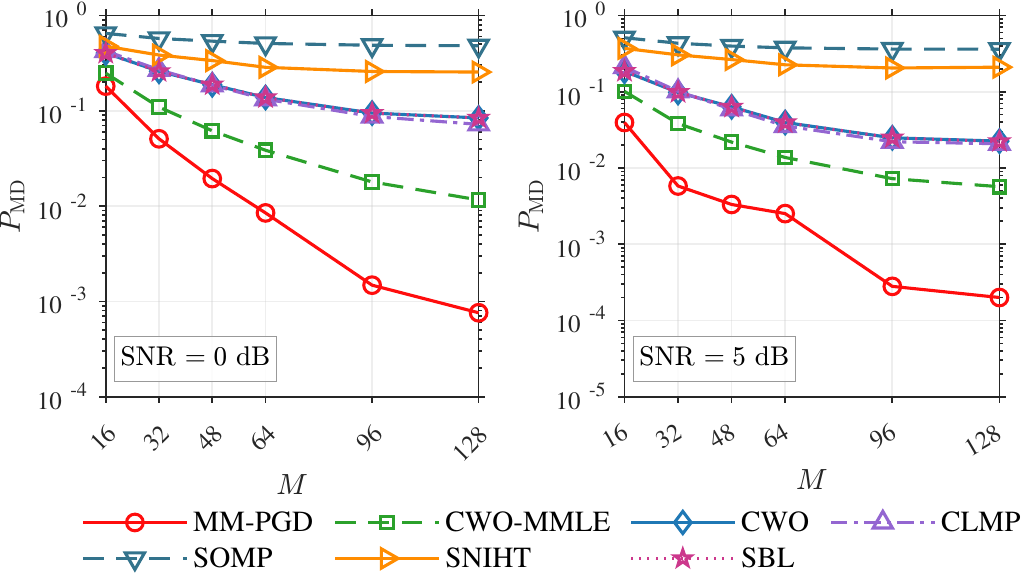}
    \caption{$P_{\mathrm{MD}}$ vs.\ number of BS antennas $M$ (Experiment~5, $R=1000$~m).
    Left: $\mathrm{SNR}=0$\,dB; Right: $\mathrm{SNR}=5$\,dB.
    $(N,K,L)=(200,25,30)$, $\kappa_{\mathrm{dB}}=-5$, $\eta_{\mathrm{NF}}=0.5$, $N_{\mathrm{MC}}=1000$. All methods benefit from a larger array, and MM-PGD gives the lowest $P_{\mathrm{MD}}$.}
    \label{fig:exp5}
\end{figure}

Fig.~\ref{fig:exp5} sweeps the antenna count $M$ from $16$ to $128$ with $R=1000$~m (enlarged from the default $500$~m to ensure $R>Z_{\mathrm{Ray}}(M)$ for all tested $M$, since $Z_{\mathrm{Ray}}\propto M^2$).
MM-PGD improves consistently with the array size: at $\mathrm{SNR}=0$\,dB, $P_{\mathrm{MD}}$ drops as the antenna count $M$ increases, and the same monotone gain is observed at $\mathrm{SNR}=5$\,dB.
All baselines benefit from a larger array, and MM-PGD gives the lowest $P_{\mathrm{MD}}$.
The widening gap with $M$ suggests that a larger array makes structured NF covariance more informative for detection: in a higher-dimensional antenna space, the per-device covariance $\bm R_n$ can capture richer spatial structure, whereas this structure is not represented under the isotropic approximation.

\subsubsection{Experiment~6: NF-user-fraction sweep}

\begin{figure}[!t]
    \centering
    \includegraphics[width=\linewidth]{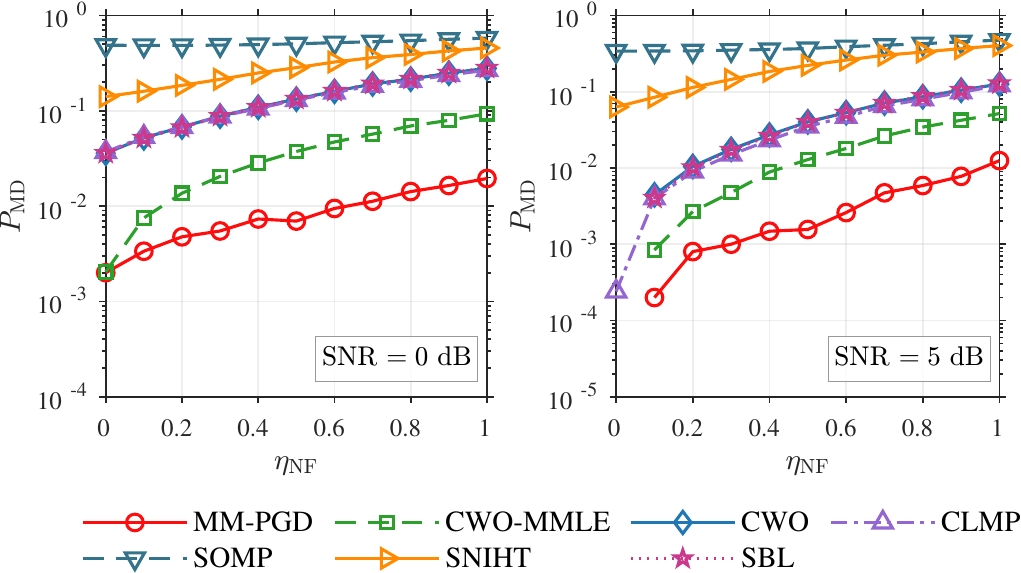}
    \caption{$P_{\mathrm{MD}}$ vs.\ NF-user fraction $\eta_{\mathrm{NF}}$ (Experiment~6).
    Left: $\mathrm{SNR} \!=\! 0$\,dB; Right: $\mathrm{SNR} \!=\! 5$\,dB.
    $(N,K,M,L)=(200,25,$ $64,30)$, $\kappa_{\mathrm{dB}} \!=\! -5$, $N_{\mathrm{MC}} \!=\! 1000$. All methods degrade as $\eta_{\mathrm{NF}}$ increases, but MM-PGD degrades the most gracefully.}
    \label{fig:exp6}
\end{figure}

Fig.~\ref{fig:exp6} sweeps the NF-user fraction $\eta_{\mathrm{NF}}$ from $0$ (all FF) to $1$ (all NF).
As $\eta_{\mathrm{NF}}$ increases, all methods degrade due to the spatially correlated near-field covariance structure.
However, MM-PGD degrades the most gracefully. At $\mathrm{SNR}=0$\,dB, its $P_{\mathrm{MD}}$ rises only from $0.0020$ in the all-FF case to $0.0195$ in the all-NF case, whereas CWO rises from $0.0359$ to $0.2789$. At $\mathrm{SNR}=5$\,dB, MM-PGD remains below $0.0125$ across the entire sweep, while CWO-MMLE stays several times higher in the high-$\eta_{\mathrm{NF}}$ region.
This relatively stable behavior of MM-PGD across $\eta_{\mathrm{NF}}$ can be understood from the unified covariance model: the gradient~\eqref{eq:mm_grad_closed} and the Kronecker--Woodbury factorization both follow the current FF/NF composition through the structure of $\{\bm C_n\}$, within the same algorithmic framework and without explicit regime-specific solver redesign.

\subsubsection{Experiment~7: Scatterer-count sweep}

\begin{figure}[!t]
    \centering
    \includegraphics[width=\linewidth]{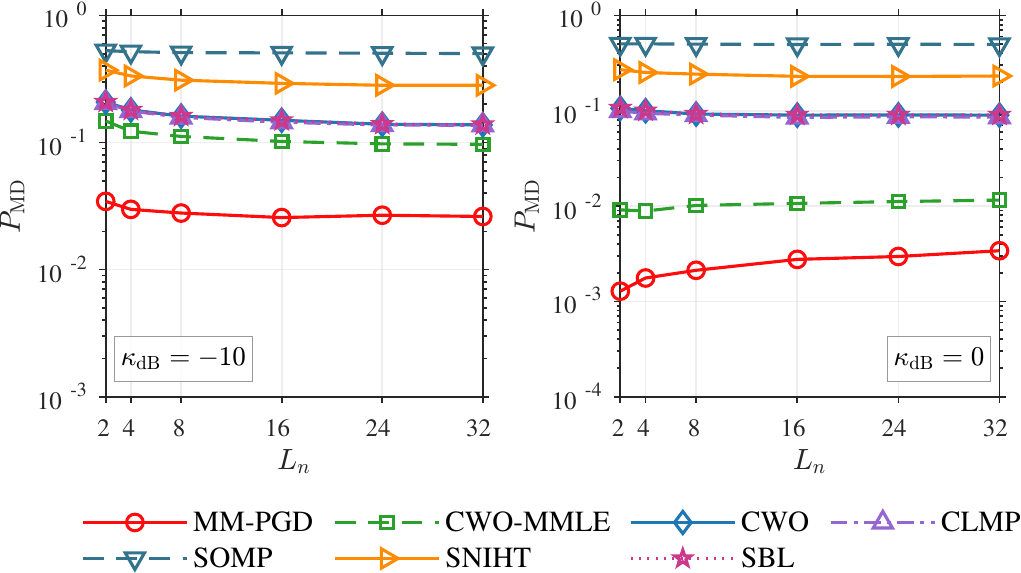}
    \caption{$P_{\mathrm{MD}}$ vs.\ number of scatterers $L_n$ (Experiment~7) at $\mathrm{SNR}=0$\,dB under two Rician regimes: $\kappa_{\mathrm{dB}}=-10$ (left) and $\kappa_{\mathrm{dB}}=0$ (right).
    $(N,K,M,L)=(200,25,64,30)$, $\eta_{\mathrm{NF}}=0.5$, $N_{\mathrm{MC}}=1000$. \revised{Larger $L_n$ helps all methods under weak LoS, but slightly degrades the mean-aware detectors (MM-PGD, CWO-MMLE) under moderate LoS.}}
    \label{fig:exp7}
\end{figure}

Fig.~\ref{fig:exp7} shows two panels at $\mathrm{SNR}=0$\,dB under weak LoS ($\kappa_{\mathrm{dB}}=-10$, left) and moderate LoS ($\kappa_{\mathrm{dB}}=0$, right).
The dominant effect across panels is the LoS strength: moving from $\kappa_{\mathrm{dB}}=-10$ to $0$ reduces MM-PGD's $P_{\mathrm{MD}}$ from the $0.026$--$0.034$ range to the $0.001$--$0.003$ range, i.e., by roughly one order of magnitude.
Within each panel, however, the scatterer count $L_n$ reveals a secondary, \emph{regime-dependent} mechanism.

Under $\kappa_{\mathrm{dB}}=-10$, the NLoS component carries roughly $91\%$ of the total channel power ($\kappa\approx 0.1$ in linear scale), making $\bm R_n$ the primary discriminative structure.
Increasing $L_n$ raises the rank of $\bm R_n$, enriching each device's spatial signature and improving separability.
All methods benefit, although MM-PGD starts from a much lower $P_{\mathrm{MD}}$ level and therefore shows a smaller but still consistent gain.

Under $\kappa_{\mathrm{dB}}=0$, a different pattern emerges. For the mean-aware detectors MM-PGD and CWO-MMLE, increasing $L_n$ no longer improves detection and may slightly degrade it. 
This is consistent with a regime in which the LoS mean already provides most of the user discrimination, so the additional NLoS covariance contributes limited extra separation. 
By contrast, covariance-only baselines can still benefit from the richer covariance structure.
Indeed, both MM-PGD and CWO-MMLE exhibit a slight $P_{\mathrm{MD}}$ increase, whereas the covariance-only baselines improve mildly as $L_n$ grows. This behavior is consistent with a regime-dependent effect of scattering richness. The scattering richness appears more beneficial when covariance is the primary information source, and less beneficial once the LoS mean becomes more dominant. %

\subsubsection{Experiment~8: Device-count sweep}

\begin{figure}[!t]
    \centering
    \includegraphics[width=\linewidth]{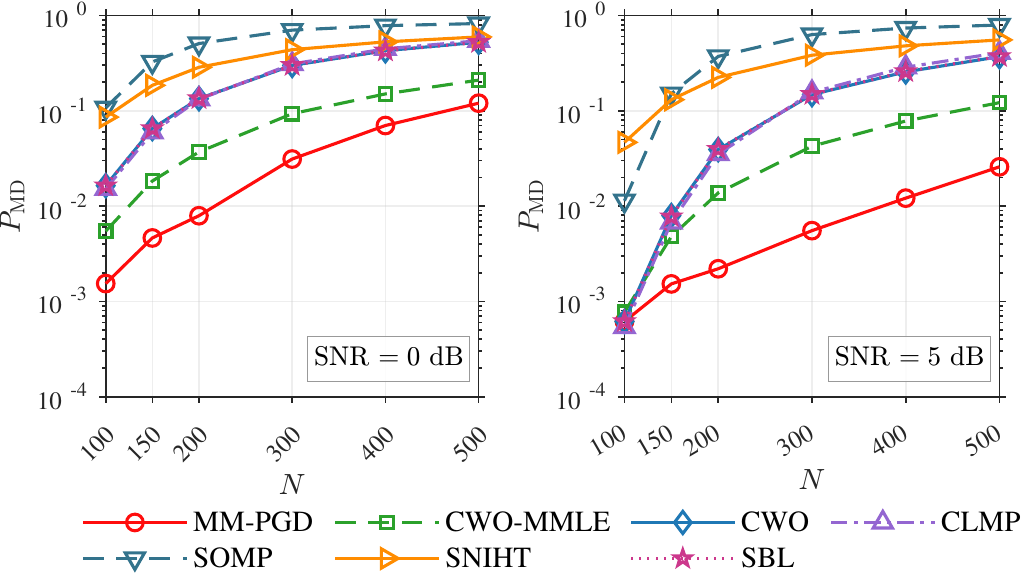}
    \caption{$P_{\mathrm{MD}}$ vs.\ total number of registered devices $N$ (Experiment~8) with fixed activity rate $K/N=12.5\%$.
    Left: $\mathrm{SNR}=0$\,dB; Right: $\mathrm{SNR}=5$\,dB.
    $(M,L)=(64,30)$, $\kappa_{\mathrm{dB}}=-5$, $\eta_{\mathrm{NF}}=0.5$, $N_{\mathrm{MC}}=1000$. As $N$ grows, MM-PGD degrades more gracefully than the other methods.}
    \label{fig:exp8}
\end{figure}

Fig.~\ref{fig:exp8} sweeps the total device count from $N=100$ to $N=500$ while keeping the activity rate fixed at $K/N=12.5\%$, so that both the pool size and the number of active devices grow together.
At $\mathrm{SNR}=5$\,dB, all methods perform comparably at the smallest scale ($N=100$), but the gap opens rapidly as $N$ grows. By $N=500$, MM-PGD reports $P_{\mathrm{MD}}=0.0258$ while CWO-MMLE degrades to $0.1217$ and CWO to $0.3698$. At $\mathrm{SNR}=0$\,dB, the same ordering holds with uniformly higher miss-detection levels.
The increasing difficulty with $N$ is expected, since larger pools introduce more inter-user interference while the pilot length $L=30$ remains fixed.
As $N$ grows, MM-PGD degrades more gracefully than the baselines. This trend is consistent with the role of structured covariance. The benefit becomes more visible as the problem scale increases. This experiment therefore complements Experiment~5 and Experiment~6 by showing that the same pattern persists under increasingly demanding user-population, array-geometry, and NF/FF-composition settings. %

\subsection{\revised{Robustness to Imperfect Statistical CSI and Non-Uniform Received Power}}
\label{subsec:sensitivity}

\revised{We assess the proposed detector under two distinct departures from the idealized setup: the first perturbs the long-term statistics supplied to the detector while leaving the true channel unchanged, and the second changes the actual per-device received power while supplying the detector with the matched statistics.}

\revised{\emph{Imperfect statistical CSI.} The system model assumes that the BS has access to the long-term statistics $(\bar{\bm h}_n,\bm R_n)$ of each device. We now assess the sensitivity of the proposed detector to imperfect statistics. The received signal $\bm Y$ is generated from the true statistics. The detector is supplied with perturbed statistics $(\widetilde{\bar{\bm h}}_n,\widetilde{\bm R}_n)$. We perturb two long-term profile parameters: the LoS angle and the large-scale path gain. The LoS angle uses $\widetilde\theta_n=\theta_n+\Delta\theta_n$, where each device's error $\Delta\theta_n\sim\mathrm{Unif}[-\Delta\theta,\Delta\theta]$ can be positive or negative, and the swept amplitude $\Delta\theta$ (the horizontal axis) runs from $0^\circ$ to $5^\circ$. The large-scale path gain uses $\widetilde\rho_n=\rho_n 10^{\xi_n/10}$, where $\xi_n\sim\mathrm{Unif}[-\Delta\rho,\Delta\rho]$ can be positive or negative and scales both moments of device $n$, and the swept amplitude $\Delta\rho$ runs from $0$ to $10$\,dB.}

\begin{figure}[!t]
    \centering
    \includegraphics[width=0.98\linewidth]{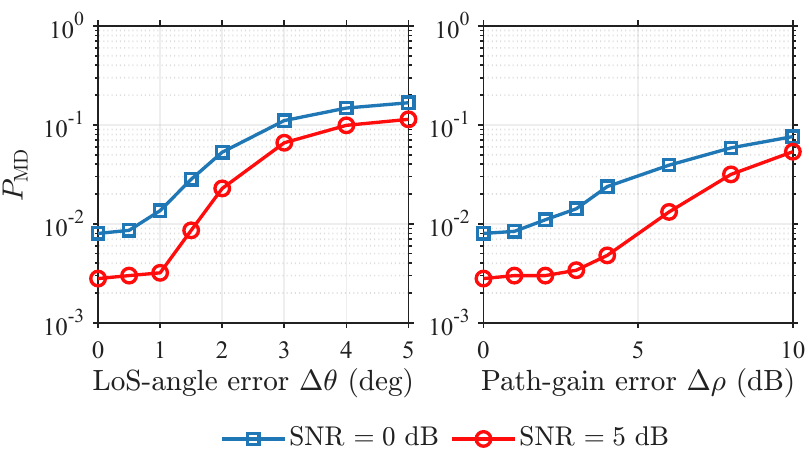}
    \caption{\revised{$P_{\mathrm{MD}}$ vs.\ LoS-angle error $\Delta\theta$ (left) and path-gain error $\Delta\rho$ (right) under imperfect statistical CSI, at $\mathrm{SNR}\in\{0,5\}$\,dB. $(N,K,M,L)=(200,25,64,30)$, $\kappa_{\mathrm{dB}}=-5$, $\eta_{\mathrm{NF}}=0.5$. The leftmost point of each curve is the perfect-prior reference.}}
    \label{fig:robust_prior}
\end{figure}

\begin{figure}[!t]
    \centering
    \includegraphics[width=0.8\linewidth]{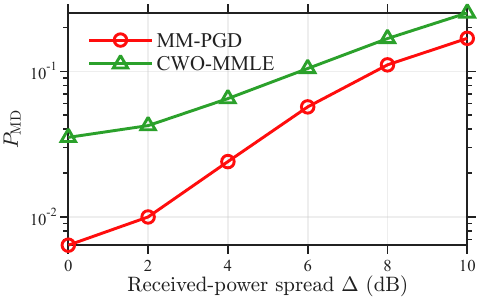}
    \caption{\revised{$P_{\mathrm{MD}}$ vs.\ the received-power spread $\Delta$, at $\eta_{\mathrm{NF}}=0.5$, $\mathrm{SNR}=0$\,dB, $(N,K,M,L)=(200,25,64,30)$, $\kappa_{\mathrm{dB}}=-5$, $N_{\mathrm{MC}}=200$. The per-device offset is $\xi_n\sim\mathrm{Unif}[-\Delta,\Delta]$\,dB, centered on the nominal received power; $\Delta=0$ is the uniform-power baseline.}}
    \label{fig:power}
\end{figure}

\begin{figure*}[!t]
    \centering
    \includegraphics[width=\textwidth]{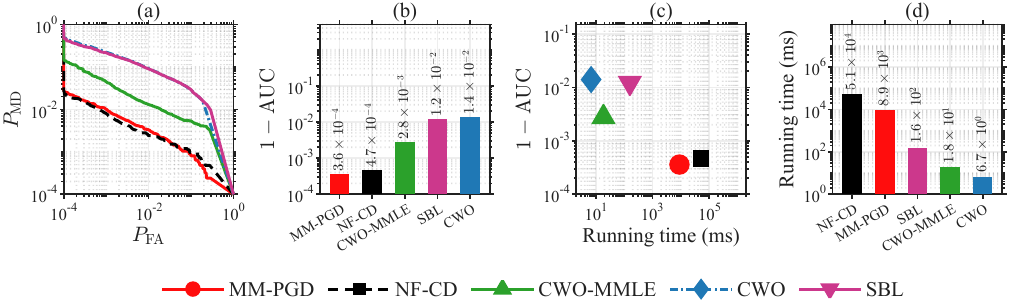}
    \caption{\revised{Threshold-based detection and accuracy--speed comparison at $L_n=8$, $\eta_{\mathrm{NF}}=0.3$, $\mathrm{SNR}=0$\,dB, $(N,K,M,L)=(200,25,64,30)$, $\kappa_{\mathrm{dB}}=-5$, $N_{\mathrm{MC}}=500$. (a)~$P_{\mathrm{MD}}$ vs.\ $P_{\mathrm{FA}}$ under a swept score threshold. (b)~$1-\mathrm{AUC}$ of algorithms (shorter is better). (c)~$1-\mathrm{AUC}$ vs.\ average per-trial running time. (d)~Running time ordered from slowest to fastest. MM-PGD and NF-CD retain the full structured NF statistics and reach the best AUC, both above the baselines, while MM-PGD is about $5.7\times$ faster than NF-CD.}}
    \label{fig:accruntime}
\end{figure*}

\revised{Figure~\ref{fig:robust_prior} presents the sensitivity of the proposed detector to the two perturbations. The LoS-angle error changes the assumed steering direction of the LoS component, and a path-gain error changes the assumed power level of each device. Both move the stored statistics away from the true channel. For small errors $P_{\mathrm{MD}}$ stays near its perfect-prior value, and for the LoS angle this plateau extends to about $1^\circ$. As the error grows further, $P_{\mathrm{MD}}$ increases smoothly. The two SNR levels follow nearly the same trend. This indicates that relying on the long-term channel statistics is practical.}

\revised{The detector relies on stored long-term statistics for every registered device. Maintaining these device-specific profiles, including the LoS mean and residual covariance, is challenging in a massive NF deployment because the statistics depend on the device geometry and scattering environment and may become outdated between refreshes. The sensitivity study shows graceful degradation under the considered LoS-angle and path-gain perturbations. Periodic offline refresh, sensing- or positioning-aided updates of the NF geometry, and mismatch-aware statistical extensions are left as future work.}

\revised{\emph{Non-uniform received power.} The preceding experiments use the common power-control normalization that scales each device's statistics so that $\mathbb E\|\bm h_n\|_2^2=\rho_0 M$, as also adopted by NF-CD~\cite{wangcas2025TWCNearField}. Since ideal power control is an idealization, we assess the detector under non-uniform received power. We draw a per-device offset $\xi_n\sim\mathrm{Unif}[-\Delta,\Delta]$ in dB centered on the nominal power, which scales the received power to $\rho_n=\rho_0 10^{\xi_n/10}$ and the stored mean and covariance to $\bar{\bm h}_n^{(\xi)}=10^{\xi_n/20}\bar{\bm h}_n$ and $\bm R_n^{(\xi)}=10^{\xi_n/10}\bm R_n$. Unlike the preceding statistical-mismatch experiment, the received signal is generated from these non-uniform statistics while the detector is given the matched ones, so the experiment isolates power heterogeneity from prior-statistics mismatch; $\Delta=0$ recovers the uniform-power baseline.}

\revised{Figure~\ref{fig:power} reports $P_{\mathrm{MD}}$ as $\Delta$ grows from $0$ to $10$\,dB. Because the per-device large-scale gain enters the model through the mean $\bar{\bm h}_n$ and the covariance $\bm R_n$, a spread of received powers is already represented in the stored statistics. The miss rate rises smoothly as the spread grows, without an abrupt change, and MM-PGD stays below the CWO-MMLE baseline over the whole range, so the detector degrades gracefully under non-uniform received power.}

\subsection{\revised{Threshold-Based Detection and Runtime Comparison}}
\label{subsec:roc}

\revised{The preceding experiments use the oracle top-$K$ rule. We now relax this rule and decide activity directly from the per-device scores. A common threshold $\tau$ is swept over its full range, so that each value of $\tau$ yields one $(P_{\mathrm{FA}},P_{\mathrm{MD}})$ pair. Figure~\ref{fig:accruntime}(a) reports the resulting $P_{\mathrm{MD}}$-versus-$P_{\mathrm{FA}}$ trade-off, the receiver operating characteristic (ROC). For any prescribed false-alarm target $\alpha$, the swept ROC can be used to read out the best operating point among the tested thresholds, e.g., $\tau^\star(\alpha)=\arg\min_{\tau:\,P_{\mathrm{FA}}(\tau)\le\alpha}P_{\mathrm{MD}}(\tau)$, with the dual interpretation for a missed-detection target. Each ROC is summarized by the area under the ROC curve (AUC), and a higher AUC is better. The AUC values here concentrate near one, and thus,  Fig.~\ref{fig:accruntime}(b) reports $1-\mathrm{AUC}$, for which lower is better. 
This evaluation also places MM-PGD against NF-CD, the strongest baseline that retains the full structured NF statistics instead of an isotropic approximation. Since NF-CD has no public implementation,  we reproduce it from the description in~\cite{wangcas2025TWCNearField} to the best of our understanding. It is given the same long-term prior as the proposed detector, namely the LoS mean and the structured low-rank NF covariance, so that the comparison isolates the algorithmic contribution from the modeling gain.}

\revised{MM-PGD and NF-CD both use the full NF statistics, and they trace essentially the same ROC curve in Fig.~\ref{fig:accruntime}(a), with AUC of $0.9996$ and $0.9995$. Figure~\ref{fig:accruntime}(b) shows the two are indistinguishable and both clearly above the isotropic CWO-MMLE, the covariance-based CWO, and SBL. The separation grows toward low $P_{\mathrm{FA}}$. The threshold-based ordering matches the ordering obtained under the oracle top-$K$ rule.}

\revised{Figure~\ref{fig:accruntime}(c) places accuracy against speed at the same operating point, plotting each detector by its $1-\mathrm{AUC}$ and its average per-trial running time. MM-PGD and NF-CD reach essentially the same AUC, and MM-PGD lies to the left of NF-CD. It attains that accuracy in $8.9$\,s against $51.0$\,s, about $5.7\times$ faster. Figure~\ref{fig:accruntime}(d) orders the wall-clock times. The isotropic-covariance and CS baselines finish in milliseconds but detect far worse. The gap follows from the update structure. MM-PGD forms all $N$ activity gradients from a fixed iterate, and hence the step parallelizes across devices, while NF-CD updates the covariance state one coordinate at a time and stays serial.}

\subsection{\revised{Convergence Behavior}}
\label{subsec:convergence}

\begin{figure}[!t]
    \centering
    \includegraphics[width=\linewidth]{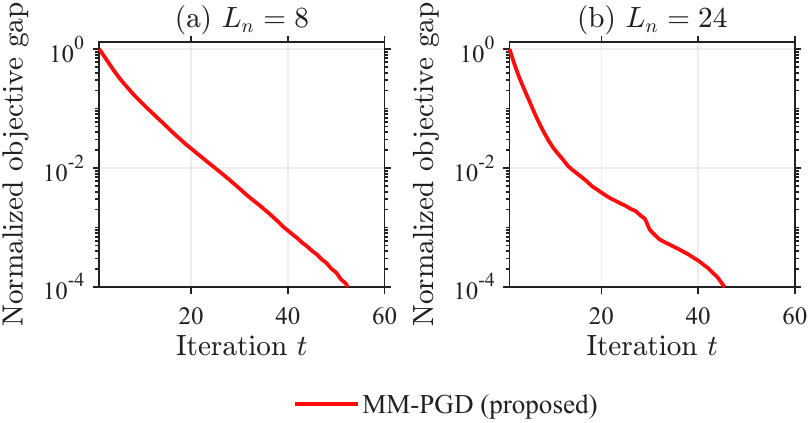}
    \caption{\revised{Normalized MM-PGD objective gap vs.\ iteration index for a representative instance at $\eta_{\mathrm{NF}}=0.5$, $\mathrm{SNR}=0$\,dB, $(N,K,M,L)=(200,25,64,30)$, $\kappa_{\mathrm{dB}}=-5$, with (a)~$8$ and (b)~$24$ scatterers per NF device.}}
    \label{fig:conviter}
\end{figure}

\revised{Proposition~\ref{prop:mm_convergence} establishes that MM-PGD generates a monotonically non-increasing objective sequence and that its projected-stationarity gap vanishes. Figure~\ref{fig:conviter} illustrates this empirically, plotting the normalized objective gap $\big(\mathcal L(\bm\gamma^{(t)})-\mathcal L^\star\big)/\big(\mathcal L(\bm\gamma^{(0)})-\mathcal L^\star\big)$ against the iteration index for a representative mixed NF/FF instance, where $\mathcal L^\star=\min_t\mathcal L(\bm\gamma^{(t)})$ is the smallest objective value attained over the run.}

\revised{The objective decreases monotonically up to numerical tolerance, dropping sharply in the first few iterations and then flattening near $\mathcal L^\star$, closing about $99\%$ of its total decrease within roughly $25$ iterations. This empirical trajectory is consistent with the monotone-descent property of Proposition~\ref{prop:mm_convergence}, and the richer near-field profile in panel~(b) follows the same trajectory.}

\subsection{Kronecker--Woodbury Runtime Validation}
\label{app:wb_benchmark}

The following experiments validate the regime-dependent computational advantage predicted by Proposition~\ref{prop:woodbury_decomp} and the complexity discussion in Section~\ref{subsec:mm_complexity}. We compare the per-trial wall-clock time of the direct Cholesky solver against the Kronecker--Woodbury solver with cross-iteration caching enabled.
Both solvers run identical MM-PGD iterations on the same random instance and return the same activity estimate; we verify numerical equivalence below.

Unless stated otherwise, the timings in Sweeps~A and~B use a fixed budget of $T_{\max}=25$ MM-PGD iterations and are averaged over $N_{\mathrm{MC}}=20$ independent trials on a single compute node (16 cores, parallel MC) with $N=200$, $K=25$, $L=30$, $\mathrm{SNR}=5$\,dB, $\kappa_{\mathrm{dB}}=-5$, $\eta_{\mathrm{NF}}=0.5$, and $R=500$\,m. The reported $r_{\mathrm{NF}}$ values refer to the effective NF rank observed in the Woodbury implementation, used only for timing interpretation; Proposition~\ref{prop:woodbury_decomp} continues to use $r_{\mathrm{NF}}$ as the total structural NF rank.

\paragraph{Sweep~A: Antenna count $M$}
We fix $L_n=4$ scatterers per NF device and sweep $M\in\{16,32,48,64,96\}$.
Since $r_{\mathrm{NF}}\approx 200$ is nearly constant across $M$ while the Cholesky cost scales as $(LM)^3$, the Woodbury advantage grows rapidly.

\begin{table}[!t]
\centering
\caption{Per-trial time (seconds) vs.\ antenna count $M$ ($L=30$, $L_n=4$).}
\label{tab:wb_sweep_a}
\begin{tabular}{r r r r r r}
\toprule
$M$ & $LM$ & Chol (s) & WB+C (s) & Speedup & $r_{\mathrm{NF}}$ \\
\midrule
16  &  480  & 16.6  & 2.0  & $8.1\times$   & 191 \\
32  &  960  & 62.5  & 3.3  & $19.1\times$  & 198 \\
48  & 1440  & 155.7 & 4.8  & $32.5\times$  & 206 \\
64  & 1920  & 335.4 & 6.4  & $52.6\times$  & 210 \\
96  & 2880  & 816.1 & 9.5  & $86.1\times$  & 213 \\
\bottomrule
\end{tabular}
\end{table}

\paragraph{Sweep~B: Scatterer count $L_n$}
We fix $M=64$ and vary the per-device scatterer count $L_n\in\{2,3,4,5,6,7,8,12,16,24,32\}$, which directly controls $r_{\mathrm{NF}}$.
The Cholesky time is roughly constant (the matrix dimension $LM=1920$ is fixed), while the Woodbury cost grows as $r_{\mathrm{NF}}$ increases.

\begin{table}[h]
\centering
\caption{Per-trial time (seconds) vs.\ scatterer count $L_n$ for $M=64$ and $L=30$.}
\label{tab:wb_sweep_b}
\begin{tabular}{r r r r r}
\toprule
$L_n$ & Chol (s) & WB+C (s) & Speedup & $r_{\mathrm{NF}}$ \\
\midrule
2   & 345.1  &   2.8  & $124.6\times$ &  113 \\
3   & 341.5  &   4.6  &  $73.7\times$ &  163 \\
4   & 340.6  &   6.5  &  $52.6\times$ &  212 \\
5   & 342.1  &   9.3  &  $36.7\times$ &  268 \\
6   & 347.8  &  13.0  &  $26.8\times$ &  328 \\
7   & 349.2  &  16.9  &  $20.7\times$ &  380 \\
8   & 346.0  &  19.7  &  $17.6\times$ &  416 \\
12  & 360.4  &  42.4  &   $8.5\times$ &  647 \\
16  & 364.2  &  70.7  &   $5.1\times$ &  827 \\
24  & 369.2  & 132.9  &   $2.8\times$ & 1182 \\
32  & 371.3  & 230.3  &   $1.6\times$ & 1519 \\
\bottomrule
\end{tabular}
\end{table}

Table~\ref{tab:wb_sweep_a} confirms the expected cubic-vs-sub-cubic scaling: the Cholesky time grows $\approx 50\times$ from $M=16$ to $M=96$, while the Woodbury time grows only $\approx 5\times$.
At the default operating point ($M=64$, $L_n=8$), the Kronecker--Woodbury solver with caching is $17.6\times$ faster; at $M=96$ the speedup reaches $86\times$.
Table~\ref{tab:wb_sweep_b} shows that the advantage diminishes as $r_{\mathrm{NF}}$ approaches $LM$, as expected from the $\mathcal O(LM\,r_{\mathrm{NF}}^2)$ complexity term.
The speedup decreases monotonically as $r_{\mathrm{NF}}/LM$ grows, remaining above $2.8\times$ at $r_{\mathrm{NF}}/LM\approx 0.62$ and falling to $1.6\times$ at $r_{\mathrm{NF}}/LM\approx 0.79$.

\paragraph{\revised{Sweep~C: Device pool size $N$}}
\revised{We fix $M=64$ and $L_n=8$ and vary the pool size $N\in\{50,100,200,400\}$ at a nominal activity ratio $K/N\approx0.125$ (so that $K=\{6,13,25,50\}$) and $\eta_{\mathrm{NF}}=0.5$, timing each $N$ over $N_{\mathrm{MC}}=5$ independent trials, so that $r_{\mathrm{NF}}\approx\eta_{\mathrm{NF}}\bar L_n N$ grows with $N$ while the full-Cholesky dimension $LM=1920$ stays fixed.}

\begin{figure}[!t]
    \centering
    \includegraphics[width=0.8\linewidth]{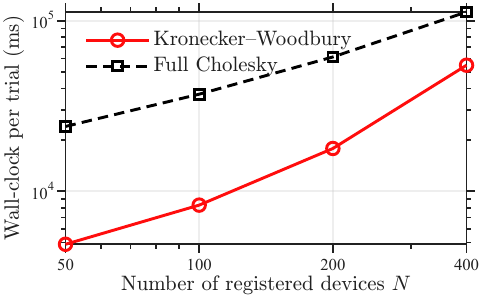}
    \caption{\revised{Average per-trial wall-clock time of the same MM-PGD detector run under the exact Kronecker--Woodbury evaluation and the direct full-Cholesky factorization, vs.\ the registered pool size $N$, at $\eta_{\mathrm{NF}}=0.5$, nominal $K/N\approx0.125$, $\mathrm{SNR}=5$\,dB, $L=30$, $M=64$ ($LM=1920$). The advantage narrows as $N$ grows, since $r_{\mathrm{NF}}$ increases toward $LM$.}}
    \label{fig:wb_nsweep}
\end{figure}

\revised{As shown in Fig.~\ref{fig:wb_nsweep}, both costs increase with $N$, but the Woodbury cost grows faster through the $\mathcal O(LM\,r_{\mathrm{NF}}^2)$ term, so the speedup narrows from about $5\times$ at $N=50$ to about $2\times$ at $N=400$. This is the empirical counterpart of the scaling estimate in Section~\ref{subsec:mm_complexity}; over the tested range $r_{\mathrm{NF}}$ stays below $LM$, so the Kronecker--Woodbury evaluation remains faster throughout.}

\paragraph{Numerical equivalence}
Across the 16 configurations of Sweeps~A and~B, the Woodbury and Cholesky solvers produce identical top-$K$ activity sets (100\% match).
The relative differences in the estimated $\bm\gamma$ vectors and NLL values are at machine precision: $\|\bm\gamma_{\mathrm{WB}}-\bm\gamma_{\mathrm{Chol}}\|/\|\bm\gamma_{\mathrm{Chol}}\|\le 10^{-9}$ and $|\mathcal L_{\mathrm{WB}}-\mathcal L_{\mathrm{Chol}}|/|\mathcal L_{\mathrm{Chol}}|\le 10^{-12}$, confirming that the decomposition is exact as stated in Proposition~\ref{prop:woodbury_decomp}.

\section{Conclusion and Discussion}
\label{sec:conclusion}
We have considered grant-free AD with mixed NF/FF users under user-specific Rician statistics in this paper. %
We developed a covariance-aware MM-PGD detector from a unified vectorized likelihood model that retains both the LoS mean and the structured mixed NF/FF covariance. 
The same update framework applies to the all-FF, all-NF, and mixed cases. 
\revised{The structural decomposition of the mixed NF/FF covariance (Proposition~\ref{prop:woodbury_decomp}) makes an exact Kronecker--Woodbury evaluation applicable, reducing the cost of covariance-matrix factorization.} %
Numerical results show that MM-PGD remains competitive in the all-FF case, while its advantage becomes more visible as the fraction of NF users increases. %
In the tested regimes, this behavior is consistent with the benefit being primarily associated with exploiting structured NF covariance.

\revised{The model assumes cross-device independence. This does not restrict the cross-antenna correlation within a device, which the structured covariance $\bm R_n$ retains and which distinguishes the mixed NF/FF setting from the antenna-domain independent-snapshot model. In dense localized deployments, nearby devices may share scattering clusters and exhibit inter-user spatial correlation, which the present per-device model does not capture. Evaluating the framework under partially shared scattering clusters is an important direction for future work.}

The proposed implementation uses an offline/online split. 
The per-device mean vectors and covariance matrices are precomputed for the registered devices, while the online detector uses only the current pilot observation and the activity variables. 
The main experiments use oracle top-$K$ support selection and assume that the long-term channel statistics $(\bar{\bm h}_n,\bm R_n)$ vary on a much slower timescale than user activity.
These assumptions isolate the detector-design problem considered here.
\revised{Our numerical study also includes a swept-threshold ROC evaluation beyond the oracle top-$K$ rule. Deployment-level adaptive threshold selection under time-varying traffic loads and long-term statistics learning remain important future directions.} %

\appendix
\subsection{Proof of Proposition~\ref{prop:gradientNLL}}
\label{app:proof_gradientNLL}
Recall $\bm r_{\bm\gamma} = \bm y-\bm\mu_{\bm\gamma}$ with
$\bm\mu_{\bm\gamma}=\sum_{k=1}^N\gamma_k\bm m_k$ and
$\bm\Sigma_{\bm\gamma}=\sum_{k=1}^N\gamma_k\bm C_k+\sigma_w^2\bm I$.
Then
$\frac{\partial \bm\Sigma_{\bm\gamma}}{\partial \gamma_n}=\bm C_n$
and
$\frac{\partial \bm r_{\bm\gamma}}{\partial \gamma_n}=-\bm m_n$.

For the log-det term, using
$\frac{\partial}{\partial x}\log|\bm\Sigma(x)|
=\tr\!\big(\bm\Sigma(x)^{-1}\frac{\partial \bm\Sigma(x)}{\partial x}\big)$,
we have
\begin{equation}
\frac{\partial \log|\bm\Sigma_{\bm\gamma}|}{\partial \gamma_n}
=\tr\!\big(\bm\Sigma_{\bm\gamma}^{-1}\bm C_n\big).
\end{equation}

For the quadratic term $\phi(\bm\gamma)\triangleq \bm r_{\bm\gamma}^\herm\bm\Sigma_{\bm\gamma}^{-1}\bm r_{\bm\gamma}$,
applying the product rule, we have
\begin{equation}
\frac{\partial \phi}{\partial \gamma_n}
=
\Big(\frac{\partial \bm r}{\partial \gamma_n}\Big)^\herm \bm\Sigma^{-1}\bm r
+\bm r^\herm \frac{\partial (\bm\Sigma^{-1})}{\partial \gamma_n}\bm r
+\bm r^\herm\bm\Sigma^{-1}\frac{\partial \bm r}{\partial \gamma_n}.
\end{equation}
Using $\frac{\partial (\bm\Sigma^{-1})}{\partial \gamma_n}
=-\bm\Sigma^{-1}\frac{\partial \bm\Sigma}{\partial \gamma_n}\bm\Sigma^{-1}
=-\bm\Sigma^{-1}\bm C_n\bm\Sigma^{-1}$
and $\bm v = \bm\Sigma_{\bm\gamma}^{-1}\bm r_{\bm\gamma}$,
we obtain
\begin{align}
\frac{\partial \phi}{\partial \gamma_n}
&= (-\bm m_n)^\herm \bm v - \bm v^\herm \bm C_n \bm v + \bm v^\herm (-\bm m_n) \nonumber \\
&= -\bm v^\herm \bm C_n \bm v - \bm m_n^\herm \bm v - \bm v^\herm \bm m_n \nonumber \\
&= -\bm v^\herm \bm C_n \bm v - 2\Re \{ \bm v^\herm \bm m_n \},
\end{align}
where the last equality holds because $\bm m_n^\herm\bm v=(\bm v^\herm\bm m_n)^\ast$.

Combining the two parts yields
\begin{equation}
\frac{\partial \mathcal L(\bm\gamma)}{\partial \gamma_n}
=
\tr(\bm\Sigma_{\bm\gamma}^{-1}\bm C_n)
-\bm v^\herm \bm C_n \bm v
-2\Re\{\bm v^\herm \bm m_n\},
\end{equation}
which proves \eqref{eq:mm_grad_closed}.

\subsection{\texorpdfstring{\revised{Proof of Proposition~\ref{prop:mm_convergence}}}{Proof of Proposition 2}}
\label{app:mm_convergence}

\revised{Let $\mathcal C=[0,1]^N$, let $\Pi_{\mathcal C}=\Pi_{[0,1]^N}$ denote the Euclidean projection onto $\mathcal C$, and let $\bm d_t=\bm\gamma^{(t+1)}-\bm\gamma^{(t)}$. The optimality condition of the projected update \eqref{eq:mm_pgd_update} is
\begin{equation}
\big(\nabla\mathcal L(\bm\gamma^{(t)})+L_t\bm d_t\big)^\top
(\bm\gamma-\bm\gamma^{(t+1)})\ge0,
\qquad \forall\bm\gamma\in\mathcal C.
\label{eq:mm_projection_optimality}
\end{equation}
Setting $\bm\gamma=\bm\gamma^{(t)}$ gives
$\nabla\mathcal L(\bm\gamma^{(t)})^\top\bm d_t\le-L_t\|\bm d_t\|_2^2$.
Combining this inequality with the acceptance condition \eqref{eq:mm_accept} yields
\begin{equation}
\mathcal L(\bm\gamma^{(t+1)})
\le
\mathcal L(\bm\gamma^{(t)})-\frac{L_t}{2}\|\bm d_t\|_2^2.
\label{eq:mm_sufficient_decrease}
\end{equation}
Because every iterate is feasible, $F(\bm\gamma^{(t)})=\mathcal L(\bm\gamma^{(t)})$. Thus, \eqref{eq:mm_sufficient_decrease} proves monotone descent, and the lower boundedness of $F$ implies convergence of the objective sequence.

The descent lemma guarantees that \eqref{eq:mm_accept} holds whenever $L_t\ge L_{\mathrm{Lip}}$. Starting from $L_0>0$, multiplying a rejected candidate by $c_{\mathrm{bt}}$, and warm-starting from the previously accepted value therefore give
$0<L_0\le L_t\le\bar L\triangleq\max\{L_0,c_{\mathrm{bt}}L_{\mathrm{Lip}}\}$.
For $G_{L_t}(\bm\gamma^{(t)})=-L_t\bm d_t$, \eqref{eq:mm_sufficient_decrease} then gives
\begin{equation}
\begin{split}
F(\bm\gamma^{(t)})-F(\bm\gamma^{(t+1)})
&\ge
\frac{1}{2L_t}\|G_{L_t}(\bm\gamma^{(t)})\|_2^2\\
&\ge
\frac{1}{2\bar L}\|G_{L_t}(\bm\gamma^{(t)})\|_2^2.
\end{split}
\label{eq:mm_mapping_decrease}
\end{equation}
Summing \eqref{eq:mm_mapping_decrease} shows that
\begin{equation*}
\sum_{t=0}^{\infty}\|G_{L_t}(\bm\gamma^{(t)})\|_2^2
\le2\bar L\big(F(\bm\gamma^{(0)})-F_{\inf}\big)<\infty.
\end{equation*}
Consequently, $G_{L_t}(\bm\gamma^{(t)})\to\bm0$, and averaging the first $T$ terms gives \eqref{eq:mm_stationarity_rate}.

Finally, let $\{\bm\gamma^{(t_j)}\}_j$ be any convergent subsequence, with $\bm\gamma^{(t_j)}\to\bm\gamma^\star$. Since $L_0\le L_{t_j}\le\bar L$, a further subsequence satisfies $L_{t_j}\to L^\star>0$. Moreover, $G_{L_{t_j}}(\bm\gamma^{(t_j)})\to\bm0$ implies $\bm\gamma^{(t_j+1)}\to\bm\gamma^\star$. Passing to the limit in \eqref{eq:mm_pgd_update} gives
\begin{equation}
\bm\gamma^\star
=
\Pi_{\mathcal C}\!\left(\bm\gamma^\star-\frac{1}{L^\star}\nabla\mathcal L(\bm\gamma^\star)\right),
\end{equation}
which, with $N_{\mathcal C}(\bm\gamma)$ denoting the normal cone of $\mathcal C$ at $\bm\gamma$, is equivalent to the stationarity inclusion
$\bm0\in\nabla\mathcal L(\bm\gamma^\star)+N_{\mathcal C}(\bm\gamma^\star)$.
Hence every accumulation point is first-order projected stationary.}

\subsection{Proof of Proposition~\ref{prop:woodbury_decomp}}
\label{app:proof_woodbury_decomp}
Partition the covariance matrix as
\begin{equation}
  \bm\Sigma_{\bm\gamma} =\sigma_w^2\bm I_{LM}+\sum_{n\in\mathcal N_{\mathrm{FF}}} \gamma_n\bm C_n +\sum_{n\in\mathcal N_{\mathrm{NF}}} \gamma_n\bm C_n.
\end{equation}
For FF devices, $\bm R_n=g_n\bm I_M$ yields
\begin{equation}
  \gamma_n\bm C_n=\gamma_n g_n\,(\bm I_M\otimes \bm s_n\bm s_n^\herm).
\end{equation}
Summing over $\mathcal N_{\mathrm{FF}}$ with the noise term gives $\bm I_M\otimes \tilde{\bm A}_{\bm\gamma}$.
For NF devices, the eigendecomposition $\bm R_n=\bm U_n\bm\Lambda_n\bm U_n^\herm$ gives
\begin{equation}
  \gamma_n\bm C_n = \sum_{k=1}^{\tilde r_n} \gamma_n\lambda_{n,k} (\bm u_{n,k}\otimes\bm s_n) (\bm u_{n,k}\otimes\bm s_n)^\herm,
\end{equation}
and concatenation yields $\tilde{\bm B}_{\bm\gamma}\tilde{\bm B}_{\bm\gamma}^\herm$.

\bibliographystyle{IEEEtran}
\bibliography{IEEEabrv,reference}

\end{document}